\documentclass[traditabstract, longauth]{aa}
\usepackage{graphicx}
\usepackage[dvips]{color} 
\usepackage{epsfig}
\usepackage{amssymb,amsmath}
\usepackage{longtable}
\usepackage{natbib}
\bibpunct{(}{)}{;}{a}{}{,} 

\begin{document}

\title{Planck pre-launch status: Low Frequency Instrument calibration and expected scientific performance}

   \author{
        A. Mennella\inst{1}\and
        M. Bersanelli\inst{1}\and
        R. C. Butler\inst{2}\and
        F.  Cuttaia\inst{2}\and
        O.  D'Arcangelo\inst{3}\and
        R. J. Davis\inst{4}\and
        M.  Frailis\inst{5}\and
        S.  Galeotta\inst{5}\and
        A.  Gregorio\inst{6}\and
        C. R.   Lawrence\inst{7}\and
        R.  Leonardi\inst{8}\and
        S. R. Lowe\inst{4}\and
        N.  Mandolesi\inst{2}\and
        M.  Maris\inst{5}\and
        P.  Meinhold\inst{8}\and
        L.  Mendes\inst{9}\and
        G.  Morgante\inst{2}\and
        M.  Sandri\inst{2}\and
        L.  Stringhetti\inst{2}\and
        L.  Terenzi\inst{2}\and
        M.  Tomasi\inst{1}\and
        L.  Valenziano\inst{2}\and
        F.  Villa\inst{2}\and
        A.  Zacchei\inst{5}\and
        A.  Zonca\inst{10}\and
        M.  Balasini\inst{11}\and
        C.  Franceschet\inst{1}\and
        P.  Battaglia\inst{11}\and
        P. M. Lapolla\inst{11}\and
        P.  Leutenegger\inst{11}\and
        M.  Miccolis\inst{11}\and
        L.  Pagan\inst{11}\and
        R.  Silvestri\inst{11}\and
            B.  Aja\inst{12}\and
            E.  Artal\inst{12}\and
            G.  Baldan\inst{11}\and
        P.  Bastia\inst{11}\and
            T.  Bernardino\inst{13}\and
            L.  Boschini\inst{11}\and
            G.  Cafagna\inst{11}\and
            B.  Cappellini\inst{10}\and
        F.  Cavaliere\inst{1}\and
            F.  Colombo\inst{11}\and
            L.  de La Fuente\inst{12}\and
            J.  Edgeley\inst{4}\and %
            M. C. Falvella\inst{14}\and 
            F.  Ferrari\inst{11}\and
            S.  Fogliani\inst{5}\and
            E.  Franceschi\inst{2}\and
            T.  Gaier\inst{7}\and
            F.  Gomez\inst{15}\and %
            J. M. Herreros\inst{15}\and
            S.  Hildebrandt\inst{15}\and
            R.  Hoyland\inst{15}\and
            N.  Hughes\inst{16}\and
            P.  Jukkala\inst{16}\and
            D.  Kettle\inst{4}\and
            M.  Laaninen\inst{17}\and
            D.  Lawson\inst{4}\and
            P.  Leahy\inst{4}\and
            S.  Levin\inst{15}\and
        P. B. Lilje\inst{18}\and
            D.  Maino\inst{1}\and
            M.  Malaspina\inst{2}\and
            P.  Manzato\inst{5}\and
        J.  Marti-Canales\inst{19}\and
            E.  Martinez-Gonzalez\inst{13}\and
            A.  Mediavilla\inst{12}\and %
            F.  Pasian\inst{5}\and
            J. P. Pascual\inst{12}\and %
            M.  Pecora\inst{11}\and
            L.  Peres-Cuevas\inst{20}\and
            P.  Platania\inst{3}\and
            M.  Pospieszalsky\inst{21}\and %
            T.  Poutanen\inst{22,23,24}\and
            R.  Rebolo\inst{16}\and
            N.  Roddis\inst{4}\and
            M.  Salmon\inst{13}\and %
            M.  Seiffert\inst{7}\and
            A.  Simonetto\inst{3}\and
            C.  Sozzi\inst{3}\and
            J.  Tauber\inst{20}\and
            J.  Tuovinen\inst{25}\and
            J.  Varis\inst{25}\and
            A.  Wilkinson\inst{4}\and
            F.  Winder\inst{4}
          }
            
   \offprints{Aniello Mennella}

   \institute{
             Universit\`a degli Studi di Milano, Dipartimento di Fisica, Via Celoria 16, 20133 Milano, Italy \and 
             INAF-IASF -- Sezione di Bologna, Via Gobetti 101, 40129 Bologna, Italy\and 
             CNR -- Istituto di Fisica del Plasma, Via Cozzi 53, 20125 Milano, Italy\and 
             Jodrell Bank Centre for Astrophysics, School of Physics \& Astronomy, University of Manchester, Manchester,
M13 9PL, U.K.\and 
             INAf -- Osservatorio Astronomico di Trieste, Via Tiepolo 11, 34143 Trieste, Italy\and 
             Universit\`a degli Studi di Trieste, Dipartimento di Fisica, Via Valerio 2, 34127 Trieste, Italy\and 
             Jet Propulsion Laboratory, California Institute of Technology, 4800 Oak Grove Drive, Pasadena, CA 91109,
USA\and 
             University of California at Santa Barbara, Physics Department, Santa Barbara CA 93106-9530, USA\and 
             Planck Science Office, European Space Agency, ESAC, P.O. box 78, 28691 Villanueva de la Caada, Madrid,
Spain\and 
             INAF-IASF -- Sezione di Milano, Via Bassini 15, 20133 Milano, Italy\and 
             Thales Alenia Space Italia, S.S Padana Superiore 290, 20090 Vimodrone (Milano), Italy\and 
             Departamento de Ingeniera de Comunicaciones, Universidad de Cantabria, Avenida De Los Castros, 39005
Santander, Spain\and 
             Instituto de Fisica De Cantabria, Consejo Superior de Investigaciones Cientificas,  
                 Universidad de Cantabria, Avenida De Los Castros, 39005 Santander, Spain\and
             Agenzia Spaziale Italiana, Viale Liegi 26, 00198 Roma, Italy\and 
             Istituto de Astrofisica de Canarias, V\'ia L\'actea, E-38200 La Laguna (Tenerife), Spain\and 
             DA-Design Oy, Keskuskatu 29, 31600 Jokioinen, Finland\and 
             Ylinen Electronics Oy, Teollisuustie 9 A, 02700 Kauniainen, Finland\and 
             Institute of Theoretical Astrophysics, University of Oslo, P.O. box 1029 Blindern, N-0315 Oslo, Norway\and
             Joint ALMA Observatory, Las Condes, Santiago, Chile\and 
             Research and Scientific Support Dpt, European Space Agency, ESTEC, Noordwijk, The Netherlands\and 
             National Radio Astronomy Observatory, 520 Edgemont Road, Charlottesville VA 22903-2475, USA\and 
             University of Helsinki, Department of Physics, P.O.Box 64 (Gustaf Hällströmin katu 2a) FI-00014 University
of Helsinki, Finland\and 
             Helsinki Institute of Physics, P.O.Box 64 (Gustaf Hällströmin katu 2a) FI-00014 University of Helsinki,
Finland\and 
             Mets\"{a}hovi Radio Observatory, Helsinki University of Technology, Metsähovintie 114 FIN-02540
Kylm\"{a}l\"{a}, Finland\and 
             MilliLab, VTT Technical Research Centre of Finland, Tietotie 3, Otaniemi, Espoo, Finland 
             }

  \date{}

  \abstract{We give the calibration and scientific performance parameters of the Planck Low Frequency Instrument (LFI)
measured during the ground cryogenic test campaign. These parameters characterise the instrument response and constitute
our best pre-launch knowledge of the LFI scientific performance.  The LFI shows excellent $1/f$ stability and rejection
of instrumental systematic effects; measured noise performance shows that LFI is  the most sensitive instrument of its
kind. The set of measured calibration parameters will be updated during flight operations through the end of the
mission.}

 
\titlerunning{LFI calibration and expected performance}
\authorrunning{A. Mennella et al}
\maketitle

%

\section{Introduction}
\label{sec:introduction}

The Low Frequency Instrument (LFI) is an array of 22~coherent differential receivers at 30, 44, and 70\,GHz on board the
European Space Agency Planck\footnote{Planck \emph{(http://www.esa.int/Planck)} is a project of the European Space
Agency - ESA - with instruments provided by two scientific Consortia funded by ESA member states (in particular the lead
countries: France and Italy) with contributions from NASA (USA), and telescope reflectors provided in a collaboration
between ESA and a scientific Consortium led and funded by Denmark.} satellite. In 15~months\footnote{There
are enough consumables on board to allow operation for an additional year.} of continuous measurements from the
Lagrangian point $L_2$ Planck will provide cosmic-variance- and foreground-limited measurements of the Cosmic Microwave
Background temperature anisotropies by scanning the sky in near great circles with a 1.5\,m dual reflector aplanatic
telescope \citep{2009_Tauber_Planck_Optics,
2004_martin_planck_telescope,2002_villa_planck_telescope,2005_dupac_planck_scanning_strategy}.

The LFI shares the focal plane of the Planck telescope with the High Frequency Instrument (HFI), an array of
52~bolometers in the 100--857\,GHz range, cooled to 0.1\,\hbox{K}. This wide frequency coverage, required for optimal
component separation, constitutes a unique feature of Planck and a formidable technological challenge, with the
integration of two different technologies with different cryogenic requirements in the same focal plane.

Excellent noise performance is obtained with receivers based on indium phosphide high electron mobility transistor
amplifiers, cryogenically cooled to 20\,K by a vibrationless hydrogen sorption cooler, which provides more than 1\,W of
cooling power at 20\,\hbox{K}. The LFI thermal design has been driven by an optimisation of receiver sensitivity and
available cooling power; in particular radio frequency (RF) amplification is divided between a 20\,K front-end unit and
a $\sim$300\,K back-end unit connected by composite waveguides \citep{2009_LFI_cal_M2}. 

The LFI has been developed following a modular approach in which the various sub-units (passive components, receiver
active components, electronics, etc.) have been built and tested individually before proceding to the next integration
step. The final integration and testing phases have been the assembly, verification, and calibration of the individual
radiometer chains~\citep{2009_LFI_cal_M4} and of the integrated instrument.

In this paper we focus on calibration, i.e., the set of parameters that provides our current best knowledge of the
instrument's scientific performance. After an overview of the  calibration philosophy we focus on the main calibration
parameters measured during test campaigns performed at instrument and satellite levels. Information concerning the test
setup and data analysis methods is provided where necessary, with references to appropriate technical articles for
further details.  The companion article that describes the LFI instrument \citep{2009_LFI_cal_M2} is the most central
reference for this paper.

The naming convention that we use for receivers and individual channels is given in
Appendix~\ref{app:naming_convention}.


\section{Overview of the LFI pseudo correlation architecture}
\label{sec:lfi_overview}

In this section we briefly summarise the LFI pseudo-correlation architecture. Further details and a more
complete treatment of the instrument can be found in \citet{2009_LFI_cal_M2}.

In the LFI each receiver couples with the Planck telescope secondary mirror via a corrugated feed horn feeding
an
orthomode transducer (OMT) that splits the incoming wave into two perpendicularly polarised components, which propagate
through two independent pseudo correlation receivers with HEMT (High Electron Mobility Transistor) amplifiers split
between a cold ($\sim$20~K) and a warm ($\sim$300~K) stage connected by composite waveguides.

\begin{figure}[h!]
    \begin{center}
       \includegraphics[width=8cm]{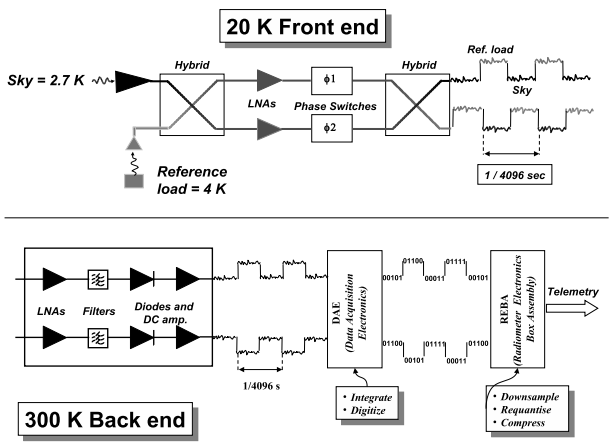}
    \end{center}
   \caption{Schematic of the LFI pseudo correlation architecture}
   \label{fig:lfi_pseudo_correlation_schematic}
\end{figure}

A schematic of the LFI pseudo correlation receiver is shown in Fig.~\ref{fig:lfi_pseudo_correlation_schematic}.
In
each radiometer connected to an OMT arm, the sky signal and the signal from a stable reference load thermally connected
to the HFI 4~K shield \citep{2009_LFI_cal_R1} are coupled to cryogenic low-noise HEMT amplifiers via a 180$^\circ$
hybrid. One of the two signals runs through a switch that applies a phase shift which oscillates between 0 and
180$^\circ$ at a frequency of 4096~Hz. A second phase switch is present for symmetry on the second radiometer 
leg but it does not introduce any phase shift. The signals are then recombined by a second 180$^\circ$ hybrid coupler,
producing a sequence of sky-load outputs alternating at twice the frequency of the phase switch.

In the back-end of each radiometer (see bottom part of
Fig.~\ref{fig:lfi_pseudo_correlation_schematic})
the RF signals are further amplified, filtered by a low-pass filter and then detected. After detection the sky and
reference load signals are integrated and digitised in 14-bit integers by the LFI DAE (Digital
Acquisition Electronics) box.

According to the scheme described above the radiometric differential power output from each diode can be written
as:

\begin{eqnarray}
   &&p_{\rm out} = a G_{\rm tot} k \beta \left[
      T_{\rm sky} + T_{\rm noise} - r \left(
      T_{\rm ref} + T_{\rm noise} \right)
      \right]\nonumber \\
   &&r = \frac{\langle V_{\rm out}^{\rm sky}\rangle}{\langle V_{\rm out}^{\rm ref}\rangle}
   \label{eq:ideal_power_output}
\end{eqnarray}
where the gain modulation factor, $r$, minimises the effect of the input signal offset between the sky ($\sim
2.7$~K)
and the reference load ($\sim 4.5$~K). The effect of reducing the offset in software and the way $r$ is estimated from
flight data are discussed in detail in \citet{mennella03}.


\section{Calibration philosophy}
\label{sec:lfi_calibration_philosophy}

The LFI calibration plan was designed to ensure optimal measurement of all parameters characterising the instrument
response. Calibration activities have been performed at various levels of integration, from single components, to the
integrated instrument, to the whole satellite. The inherent redundancy of this approach provided maximum knowledge of
the instrument and of its sub-units, as well as calibration at different levels.

Table~\ref{tab:list_main_calibration_parameters} gives the main LFI instrument parameters and the integration levels at
which they have been measured. Three main groups of calibration activities are identified: (i) basic calibration
(Sect.~\ref{sec:basic_calibration}), (ii) receiver noise properties (Sect.~\ref{sec:noise}), and (iii) susceptibility
(Sect.~\ref{sec:susceptibility}).

A particular point must be made about the front-end bias tuning, which is not part of calibration but is nevertheless a
key step in setting the instrument scientific performance. In order to satisfy tight mass and power constraints, power
bias lines have been divided into four common-grounded power groups, with no bias voltage readouts. Only the total drain
current flowing through the front-end amplifiers is measured and is available in the housekeeping telemetry.  This
design has important implications on front-end bias tuning, which depends critically on the satellite electrical and
thermal configuration. Therefore front-end bias tuning has been repeated at all integration stages, and will also be
repeated in flight before the start of nominal operations. Details about bias tuning performed at the various
integration levels can be found in \citet{2009_LFI_cal_R8}, \citet{2009_LFI_cal_R10}, \citet{2009_LFI_cal_M4} and
\citet{2009_LFI_cal_R7}.

\begin{table*}
   \caption{
        Main instrument parameters and stages at which they have been / will be measured. In bodface we highlight
calibration parameters defining the instrument scientific performance that are discussed in this paper.
    }
   \label{tab:list_main_calibration_parameters}
   \begin{small}
   \begin{center}
      \begin{tabular}{|p{2.5cm}|p{2.5cm}|p{2.5cm}|p{2cm}|p{2cm}|p{2cm}|p{2cm}|}
         \hline
         \vspace{.025cm}
         \textbf{Category} & \vspace{.025cm}\textbf{Parameters} &
         \vspace{.025cm}\textbf{\centerline{Additional}} & 
         \vspace{.025cm}\textbf{\centerline{Individual}} & 
         \vspace{.025cm}\textbf{\centerline{Integrated}} & 
         \vspace{.025cm}\textbf{\centerline{Satellite}} & 
         \vspace{.025cm}\textbf{\centerline{In flight}} \\
         
         &  
         & \textbf{\vspace{0.025cm}\centerline{Reference}} 
         & \textbf{\vspace{0.025cm}\centerline{radiometers}} 
         & \textbf{\vspace{0.025cm}\centerline{instrument}} 
         &  
         &  \\
         \hline
         
         \vspace{.025cm}\textbf{\textit{Bias tuning}} & 
         \vspace{.025cm}Front-end amplifiers &
         \vspace{.025cm}\citet{2009_LFI_cal_R7} &  
         \vspace{0.025cm}\vspace{0.025cm}\vspace{0.025cm}\centerline{Y} & 
         \vspace{0.025cm}\centerline{Y} & \vspace{0.025cm}\centerline{Y} & 
         \vspace{0.025cm}\centerline{Y} \\
         \cline{2-7} & 
         
         \vspace{.025cm}Phase switches & 
         \vspace{.025cm}\citet{2009_LFI_cal_R7} &   
         \vspace{0.025cm}\centerline{Y} & 
         \vspace{0.025cm}\centerline{Y} & 
         \vspace{0.025cm}\centerline{Y} & 
         \vspace{0.025cm}\centerline{Y} \\
         \hline
         \multicolumn{7}{|c|}{\textbf{Calibration}}\\
         \hline
         \vspace{.025cm}\textbf{\textit{Basic calibration}}  & 
         \vspace{.025cm}\textbf{Photometric calibration} & 
         \vspace{0.025cm} \citet{2009_LFI_cal_M4}&
         \vspace{0.025cm}\centerline{Y} & 
         \vspace{0.025cm}\centerline{Y} & 
         \vspace{0.025cm}\centerline{Y} & 
         \vspace{0.025cm}\centerline{Y} \\
         
         \cline{2-7} & 
         \vspace{.025cm}\textbf{Linearity} & 
         \vspace{.025cm}\citet{2009_LFI_cal_R4} &
         \vspace{0.025cm}\centerline{Y} & 
         \vspace{0.025cm}\centerline{Y} & 
         \vspace{0.025cm}\centerline{N} & 
         \vspace{0.025cm}\centerline{N} \\
         
         \cline{2-7} & 
         \vspace{.025cm}\textbf{Isolation} & 
         \vspace{0.025cm} \citet{2009_LFI_cal_M4}&
         \vspace{0.025cm}\centerline{Y} & 
         \vspace{0.025cm}\centerline{Y} & 
         \vspace{0.025cm}\centerline{N} & 
         \vspace{0.025cm}\centerline{N} \\
         
         \cline{2-7} & 
         \vspace{.025cm}In-band response & 
         \vspace{.025cm}\citet{2009_LFI_cal_R3} &
         \vspace{0.025cm}\centerline{Y} & 
         \vspace{0.025cm}\centerline{N} & 
         \vspace{0.025cm}\centerline{N} & 
         \vspace{0.025cm}\centerline{N} \\
         
         \hline
         \vspace{.025cm}\textbf{\textit{Noise performance}}  & 
         \vspace{.025cm}\textbf{White noise} & 
         \vspace{.025cm}\citet{2009_LFI_cal_R2} &
         \vspace{0.025cm}\centerline{Y} & 
         \vspace{0.025cm}\centerline{Y} & 
         \vspace{0.025cm}\centerline{Y} & 
         \vspace{0.025cm}\centerline{Y} \\
         
         \cline{2-7} & 
         \vspace{.025cm}\textbf{Knee frequency} & 
         \vspace{.025cm}\citet{2009_LFI_cal_R2} &
         \vspace{0.025cm}\centerline{Y} & 
         \vspace{0.025cm}\centerline{Y} & 
         \vspace{0.025cm}\centerline{Y} & 
         \vspace{0.025cm}\centerline{Y} \\
               
         \cline{2-7} & 
         \vspace{.025cm}\textbf{1/$f$ slope} & 
         \vspace{.025cm}\citet{2009_LFI_cal_R2} &
         \vspace{0.025cm}\centerline{Y} & 
         \vspace{0.025cm}\centerline{Y} & 
         \vspace{0.025cm}\centerline{Y} & 
         \vspace{0.025cm}\centerline{Y} \\
         
         \hline
         \vspace{.025cm}\textbf{\textit{Susceptibility}} & 
         \vspace{.025cm}\textbf{Front-end temperature fluctuations} & 
         \vspace{.025cm}\citet{2009_LFI_cal_R6} &
         \vspace{0.025cm}\centerline{Y} & 
         \vspace{0.025cm}\centerline{Y} & 
         \vspace{0.025cm}\centerline{Y} & 
         \vspace{0.025cm}\centerline{Y} \\
         
         \cline{2-7} & 
         \vspace{.025cm}Back-end temperature fluctuations & 
          &
         \vspace{0.025cm}\centerline{Y} & 
         \vspace{0.025cm}\centerline{Y} & 
         \vspace{0.025cm}\centerline{N} & 
         \vspace{0.025cm}\centerline{N} \\
               
         \cline{2-7} & 
         \vspace{.025cm}Front-end bias fluctuations & 
          &
         \vspace{0.025cm}\centerline{Y} & 
         \vspace{0.025cm}\centerline{Y} & 
         \vspace{0.025cm}\centerline{N} & 
         \vspace{0.025cm}\centerline{N} \\
         \hline
      \end{tabular}
   \end{center}
   \end{small}
\end{table*}


\section{Instrument-level cryogenic environment and test setup}
\label{sec:cryo_and_setup}

The LFI receivers and the integrated instrument were tested in 2006 at the Thales Alenia Space-Italia laboratories
located in Vimodrone (Milano). Custom-designed cryo-facilities were developed in order to reproduce as closely as
possible flight-like thermal, electrical, and data interface conditions \citep{2009_LFI_cal_T1}.
Table~\ref{tab:summary_flight_testing_conditions} compares the main expected flight thermal conditions with those
reproduced during tests on individual receivers and on the integrated instrument.

\begin{table}[h!]
   \caption{
      Summary of main thermal conditions in flight and in the various testing facilities.
      }
   \label{tab:summary_flight_testing_conditions}
   \begin{center}
      \begin{tabular}{|l|c|c|c|}
         \hline
          Temperatures            &  Flight & Receiver    & Instr.\\
         \hline
         Sky &  $\sim$ 3~K & $\gtrsim$ 8~K & $\gtrsim$ 18.5~K \\
         \hline
         Ref. & $\sim$ 4.5~K & $\gtrsim$ 8~K & $\gtrsim$ 18.5~K \\
         \hline
         Front-end& $\sim$ 20~K & $\sim$ 20~K & $\sim$ 26~K  \\
         \hline
         Back-end & $\sim$ 300~K & $\sim$ 300~K & $\sim$ 300~K \\
         \hline
      \end{tabular}
   \end{center}
\end{table}

As can be seen from the table, during the integrated instrument tests the temperature of the sky and reference loads was
much higher than expected in flight (18.5\,K vs.~3--4.5\,K).  To compensate for this, receiver-level tests were
conducted with the sky and reference loads at two temperatures, one near flight, the other near 20\,K
\citep{2009_LFI_cal_M4}.  During the instrument-level tests, parameters dependent on the sky and reference load
temperatures (such as the white noise sensitivity and the photometric calibration constant) could be extrapolated to
flight conditions.

\subsection{Thermal setup}
\label{sec:thermal_setup}

\begin{figure}[h!]
         \begin{center}
             \includegraphics[width=8cm]{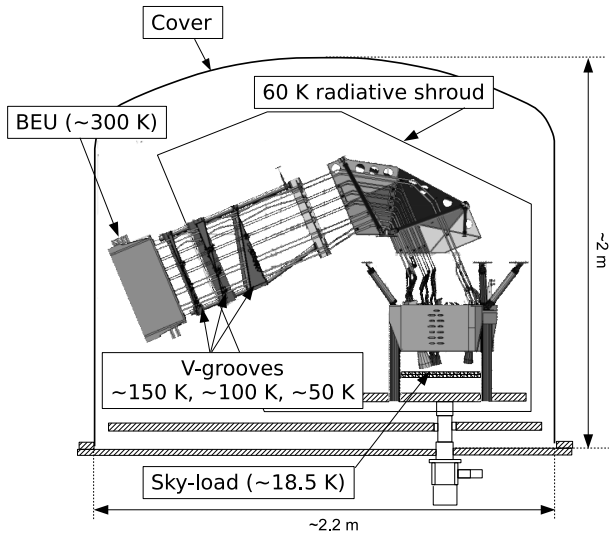}
         \end{center}
        \caption{
            LFI cryo-chamber facility. The LFI is mounted face-down with the feed horn array facing the eccosorb
sky-load
        }
        \label{fig:lfi_cryochamber}
   \end{figure}

   A schematic of the LFI cryo-facility with the main thermal interfaces is shown in Fig.~\ref{fig:lfi_cryochamber}. The
LFI was installed face-down, with the feed-horns directed towards an ECCOSORB ``sky-load'' and the back-end unit resting
upon a tilted support.  The entire instrument was held in place by a counterweight system that allowed slight movements
to compensate for thermal contractions during cooldown. The reference loads were mounted on a mechanical structure
reproducing the HFI external interfaces inserted in the middle of the front-end unit. 

   We summarise here and in Tab.~\ref{tab:summary_cryofacility_thermal_performance} the main characteristics and issues
   of the testing environment. Further details about the sky load thermal design can be found in
\citet{2009_LFI_cal_T1}. 

    \textbf{Front-end unit}. The front end unit and the LFI main frame were cooled by a large copper flange simulating
the sorption cooler cold end interface. The flange was linked to the 20\,K cooler by means of ten large copper braids. 
Its temperature was controlled by a PID controller, and was stable to $\sim$35\,mK at temperatures $\gtrsim 25.5$\,K at
the control stage. The thermal control system was also used in the susceptibility test to change the temperature of the
front end in steps (see Sect~\ref{sec:susceptibility}).
      
     \textbf{Sky load}. The sky load was thermally linked to the 20\,K cooler through a gas heat switch that could be
adjusted to obtain the necessary temperature steps during calibration tests. One of the sensors mounted in the central
region of the load did not work correctly during the tests and results from the thermal modelling were used to detail
its thermal behaviour.

    \textbf{Reference loads}. The reference loads were installed on an aluminium structure thermally anchored to the
20~K cooler by means of high conductivity straps. 

    An upper plate held all 70\,GHz loads, while the 30 and 44\,GHz loads were attached to three individual flanges. Two
thermometers on the bottom flange were used to measure and control the temperature of the whole structure. Five other
sensors monitored the temperatures of the aluminum cases of the reference loads.  The average temperature of the loads
was around 22.1\,K, with typical peak-to-peak stability of 80\,mK. 

    \textbf{Radiative shroud}. The LFI was enclosed in a thermal shield intercepting parasitics and providing a
cold radiative environment.  The outer surface was highly reflective, while the inner surface was coated black to
maximise radiative coupling. Two 50\,K refrigerators cooled the thermal shield to temperatures in the range 43--70\,K, 
depending on the distance to the cryocooler cold head, as measured by twelve diode sensors.
 
    \textbf{Back-end unit}. The warm back-end unit was connected to a water circuit with temperature stabilised
by a PID (Proportional, Integral, Derivative) controller; this stage suffered from diurnal temperature instabilities of
the order of $\sim$0.5\,K peak-to-peak. The effect of these temperature instabilities was visible in the total power
voltage output from some detectors, but was almost completely removed by differencing.
 
   \begin{table}[h!]
      \caption{Summary of the main LFI cryo-facility thermal performance. The temperature stability listed in
the second column refers to the measured peak-to-peak during one day.}
      \label{tab:summary_cryofacility_thermal_performance}
      \begin{center}
         \begin{tabular}{l c c}
            \hline
            \hline
                        &Avg. Temp. (K) &Stability (K)\\
            \hline
                    Sky load            &18--35 &0.10\\
                    Focal plane unit    &26     &0.03\\
                    Reference loads     &22     &0.08\\
                    Back end unit       &315    &0.65 \\
            \hline
         \end{tabular}

      \end{center}

   \end{table}


\section{Measured calibration parameters and scientific performance}
\label{sec:results_overview}

In this section we present the main calibration and performance parameters (refer to 
Table~\ref{tab:list_main_calibration_parameters}).

During the instrument-level test campaign we experienced two failures, one on the 70\,GHz radiometer \texttt{LFI18M-0}
and the other on the 44\,GHz radiometer \texttt{LFI24M-0}. The \texttt{LFI18M-0} failure was caused by a phase switch
that cracked during cooldown. At the end of the test campaign, just before instrument delivery to ESA,
the radiometer LFI18M-0 was replaced with a flight spare. In the second case the problem was a
defective electrical contact to the amplifier $V_{\rm g2}$ (gate 2 voltage) line, which was repaired after the end of
the test. Subsequent room-temperature tests as well cryogenic ground satellite tests (Summer 2008) and in-flight
calibration (Summer 2009) showed full functionality, confirming the successful repair of \texttt{LFI18M-0} and
\texttt{LFI24M-0}. Because these two radiometers were in a failed state during the test campaign, we generally show no
results from them. The only exception is the calibrated noise per frequency channel reported in
Table~\ref{tab:calibrated_white_noise_results}, where:

\begin{itemize}
    \item for \texttt{LFI18M-0}, we assume the same noise parameters obtained for \texttt{LFI18S-1}; and 
    \item for \texttt{LFI24M-0}, we use the noise parameters measured during single-receiver tests before integration
into the instrument array.
\end{itemize}
   
\subsection{Basic calibration}
\label{sec:basic_calibration}

    \subsubsection{Experimental setup}
    \label{sec:calibration_noise_temperature_experimental}
   
        These parameters have been determined by means of tests in which the radiometric average voltage output, $V_{\rm
out}$, was recorded for various input antenna temperature levels, $T_{\rm in}$. Although straightforward in principle,
these tests required the following conditions in the experimental setup and in the measurement procedure in order to
maximise the achieved accuracy in the recovered parameters:
   
        \begin{itemize}

                \item the sky load temperature distribution must be accurately known;

                \item temperature steps must be large enough (at least few Kelvin) to dominate over variations 
                      caused by $1/f$ noise or other instabilities;

                \item the reference load temperature must remain stable during the change in the sky load temperature
or, alternatively, variations must be taken into account in the data analysis especially in the determination of
receiver isolation;     

                \item data points must be acquired at multiple input temperatures to increase accuracy in estimating
response linearity.       

        \end{itemize}
    
        These conditions were all met during receiver-level tests in which several steps were obtained over a
temperature span ranging from $\sim$8~K to $\sim$30~K and where the sky-load temperature distrubution was very well
known both experimentally and from thermal modelling \citep{2009_LFI_cal_T4,2009_LFI_cal_M4}. 
        
        On the other hand, during instrument-level tests these conditions were not as well-met:
    
        \begin{itemize}
                \item the total number of available temperature controllers allowed us to place only three sensors on
the sky load, one on the back metal plate, one on the side, and one on the tip of the central pyramid. The input
temperature was then determined using the measurements from these three sensors in a dedicated thermal model of the sky
load itself;    
            
                \item the minimum and maximum temperatures that could be set without impacting the focal plane and
reference load temperatures were 17.5\,K and 30\,K, half the range obtained during receiver-level tests;     
            
                \item the time needed to change the sky load temperature few kelvin was large, of the order of several
hours, because of its large thermal mass. This limited to three the number of temperature steps that could be performed
in the available time.   
        \end{itemize}
    
        The reduced temperature range and number of discrete temperatures that could be set precluded determination of
the linearity factor. which was therefore excluded from the fit and constrained to $\pm 1$\% around the value found
during calibration of individual receivers (see Sect.~\ref{sec:calibration_noise_temperature}).\footnote{The slight
compression found in the output of the 30 and 44\,GHz receivers is caused by the back-end amplifier and diode, which
worked at the same conditions both during both test campaigns.}
    
        Table~\ref{tab:basic_calibration_temperatures} summarises temperatures for the three temperature steps
considered in these tests. The sky load temperature (antenna temperature), has been determined from the sky load thermal
model using temperature sensor data. The reference load temperature is a direct measurement converted into antenna
temperature. Front-end and back-end unit temperatures are direct temperature sensor measurements averaged over all
sensors.

        \begin{table}[h!]
            \caption{Main temperatures during basic calibration temperature steps. }
            \label{tab:basic_calibration_temperatures}
            \begin{center}
                \begin{tabular}{c c c c c}
                \hline  
                \hline
                Step \# &  $T_{\rm sky}$(K) & $T_{\rm ref}$ (K) & $T_{\rm FEU}$ (K)& $T_{\rm BEU}$ (C)\\
                \hline
                1       &   22.05        &   22.34       &    26.40      &    37.53      \\
                2       &   28.96        &   22.20       &    26.45      &    37.48      \\
                3       &   32.91        &   22.32       &    26.40      &    37.67      \\
                \hline
                \end{tabular}
            \end{center}
    
        \end{table}
    
    \subsubsection{Photometric calibration, noise temperature and linearity}
    \label{sec:calibration_noise_temperature}
    
        Noise temperatures and calibration constants can be calculated by fitting the $V_{\rm out}(T_{\rm sky})$ data
with the most representative model \citep{1989_daywitt_nonlinear_equations, 2009_LFI_cal_R4}:
    
        \begin{equation}
                V_{\rm out} = \frac{G_0(T_{\rm sky} + T_{\rm noise})}{1 + b\, G_0(T_{\rm sky} + T_{\rm noise})}
                \label{eq:nonlinear_response}
        \end{equation}
        where $V_{\rm out}$ is the voltage output, $T_{\rm sky}$ is the sky load input antenna temperature, $T_{\rm
noise}$ is the noise temperature, $G_0$ is the photometric calibration constant in the limit of linear response, and $b$
is the nonlinearity parameter. For perfectly linear receivers $b=0$.
    
        In Table~\ref{tab:tnoise_gain_lin_results} we summarise the best-fit parameters obtained for all the LFI
detectors. The nonlinearity parameter $b$ for the 70\,GHz receivers is $\lesssim 10^{-3}$, consistent with zero within
the measurement uncertainty. The 30 and 44\,GHz receivers show some compression at high input temperatures. This
nonlinearity arises from the back-end RF amplification stage and detector diode, which show compression down to very low
input powers. The nonlinear response has been thoroughly tested both on the individual back end modules
\citep{2009_LFI_cal_R4} and during the RCA calibration campaign \citep{2009_LFI_cal_M4} and has been shown to fit
well Eq.~(\ref{eq:nonlinear_response}).
    
    \subsubsection{Isolation}
    \label{sec:isolation}
    
        Isolation was estimated from the average radiometer voltage outputs, $V_{\rm sky}$ and $V_{\rm ref}$, at the two
extreme sky load temperatures (Steps 1 and 3 in Table~\ref{tab:basic_calibration_temperatures})\footnote{The test can be
conducted, in principle, also by changing the reference load temperature. In the instrument cryofacility, however, this
was not possible because only the sky load temperature could be controlled.}. Equations used to calculate isolation
values and uncertainties are reported in Appendix~\ref{app:isolation}.
    
        In Fig.~\ref{fig:isolation_results} we summarise the measured isolation for all detectors and provide a
comparison with similar measurements performed on individual receiver chains. The results show large uncertainties in
isolation measured during instrument-level tests, caused by $1/f$ noise instabilities in the total power datastreams
that were not negligible in the time span between the various temperature steps that was of the order of few days.
    
        \begin{figure}[h!]
             \begin{center}
             \includegraphics[width = 4.3cm]{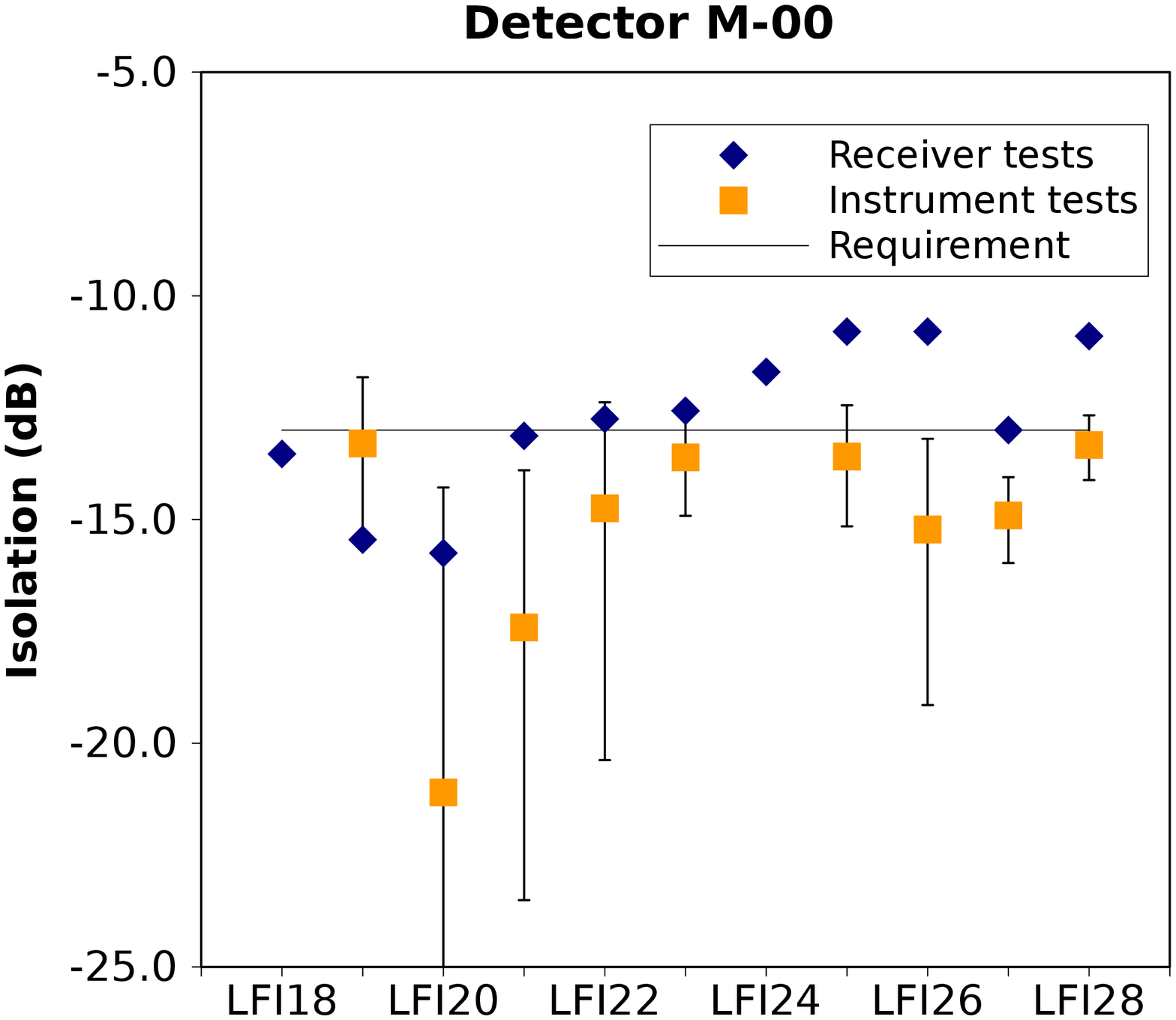} 
             \includegraphics[width = 4.3cm]{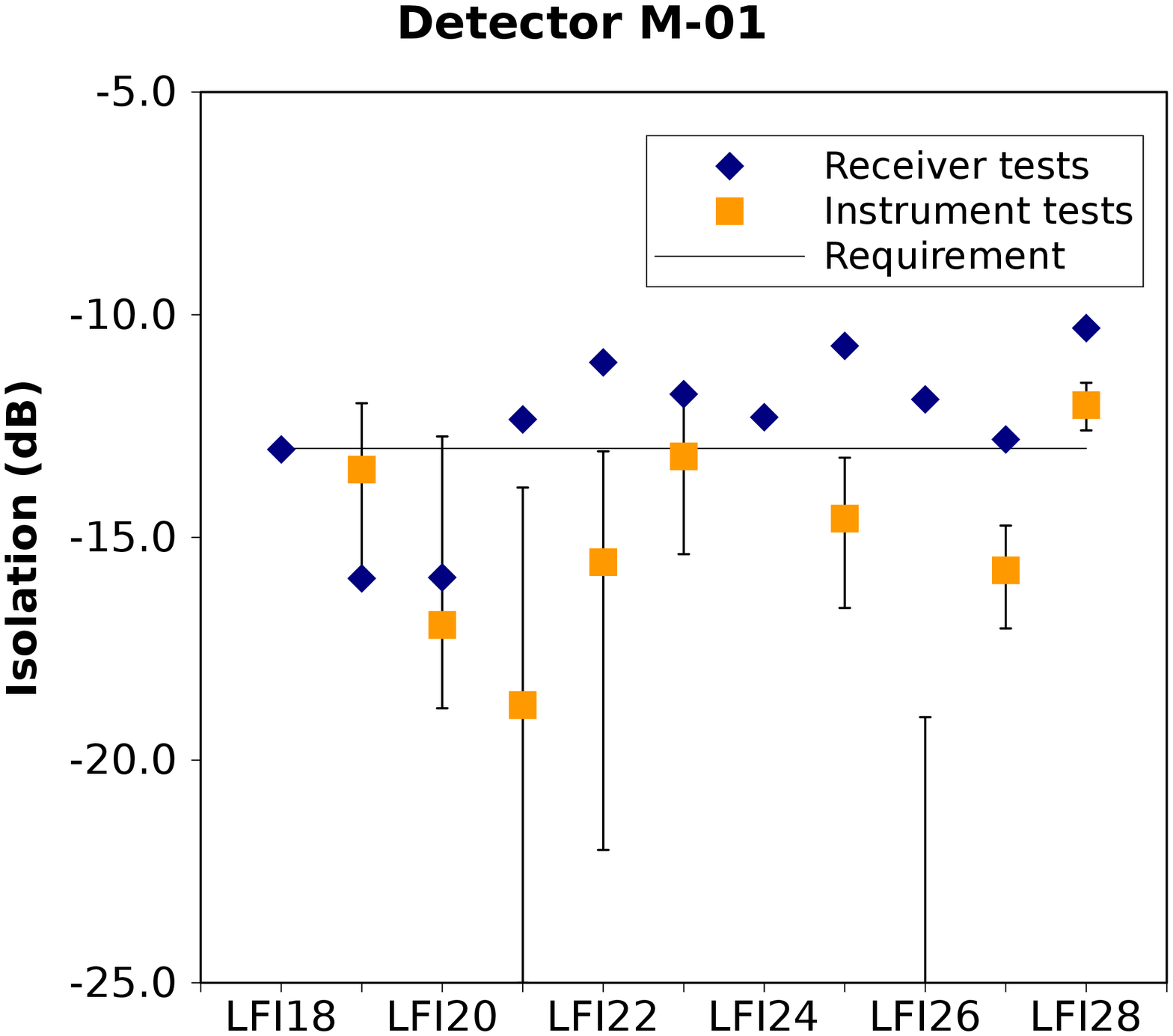} \\
             \vspace{0.5cm}
             \includegraphics[width = 4.3cm]{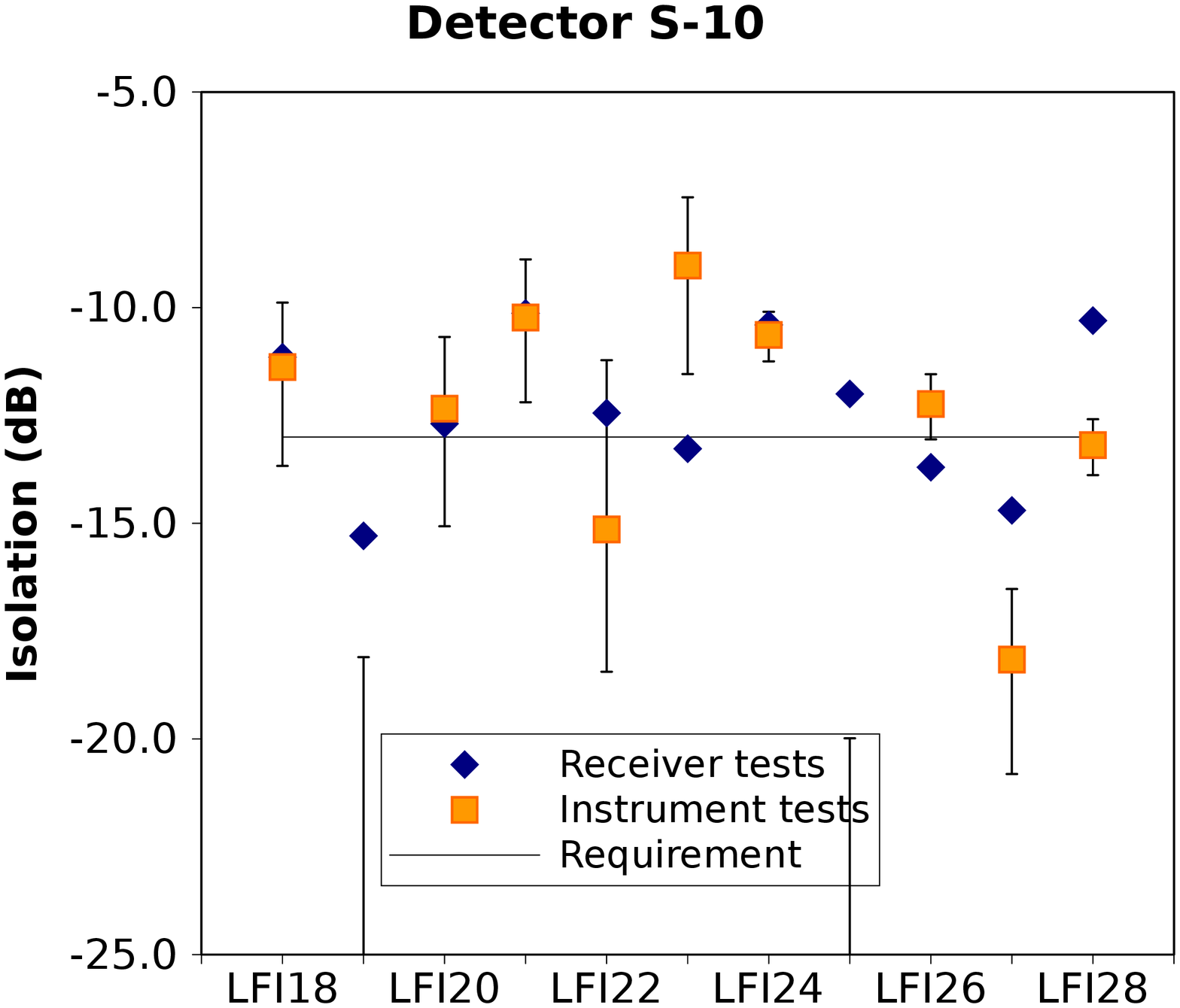} 
             \includegraphics[width = 4.3cm]{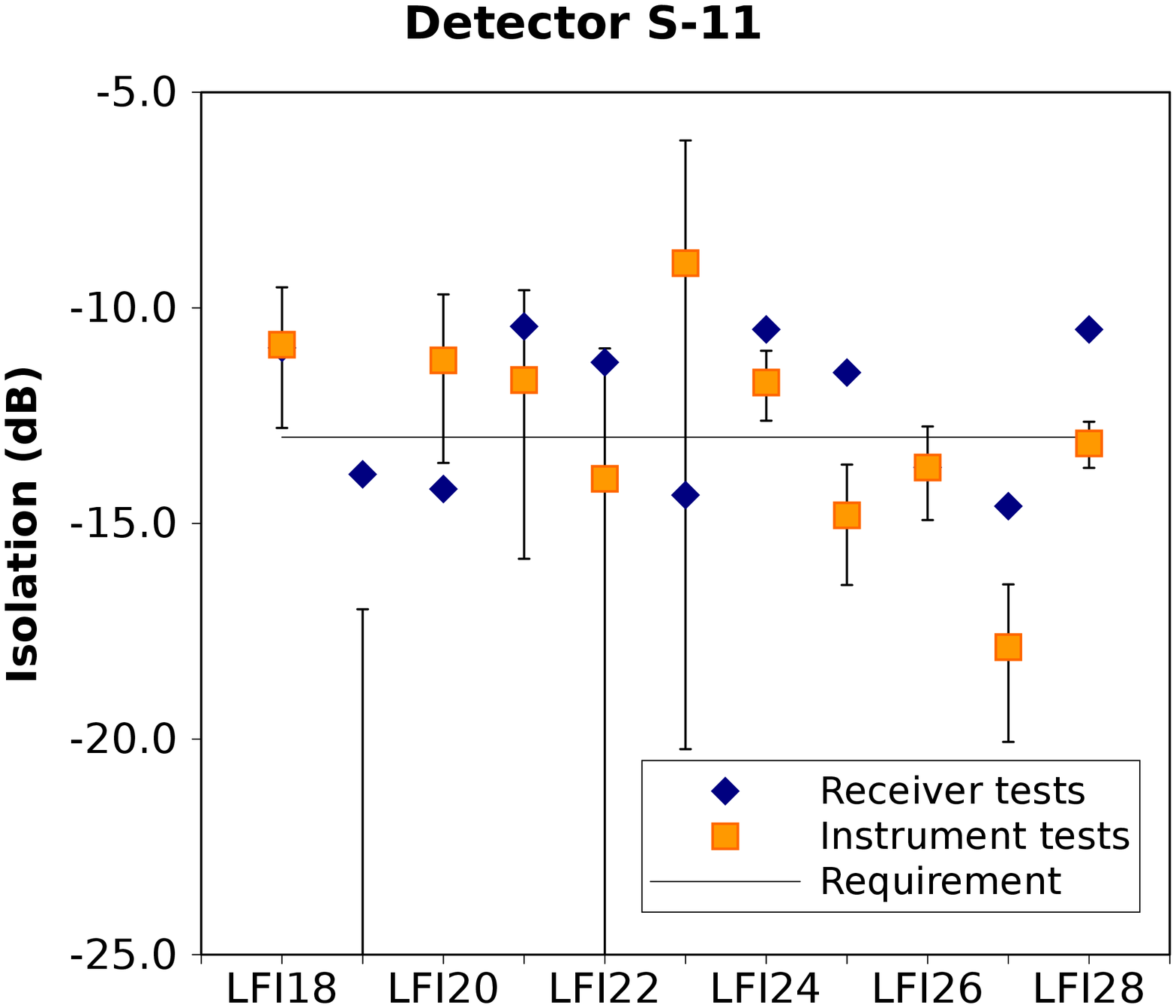} 
             \end{center}
    
            \caption{
            Summary of measured isolation compared with the same measurements
            performed at receiver level \citep{2009_LFI_cal_M4}.
            }
            \label{fig:isolation_results}
        \end{figure}
    
        Apart from the limitations given by the measurement setup, the results show that isolation lies in the range
$-10$\,dB to $-20$\,dB, which is globally within the  requirement of $-13$~dB.


\subsection{Noise properties}
\label{sec:noise}

    The pseudo-correlation design of the Planck-LFI receivers has been optimised to minimise the effects of $1/f$ gain
    variations in the radiometers. 
    
    The white noise sensitivity of the receivers is essentially independent of the reference load temperature level
    \citep{seiffert02} and can be written, in its most general form, as follows:
    
    \begin{equation}
        \Delta T_{\rm rms} = K \frac{T_{\rm sky}+T_{\rm noise}}{\sqrt{\beta}}
        \label{eq:white_noise_sensitivity}
    \end{equation}
    where $\beta$ is the receiver bandwidth, $\Delta T_{\rm rms}$ is the white noise sensitivity per unit integration
    time, and $K$ is a constant. 

    For data obtained from a single diode output, $K=1$ for unswitched data and $K=2$ for differenced data.  The factor
    of 2 for differenced data is the product of one $\sqrt{2}$ from the difference and another $\sqrt{2}$ from the
    halving of the sky integration time. When we average the two (calibrated) outputs of each radiometer we gain
    back a factor $\sqrt{2}$, so that the final radiometer sensitivity is given
    by Eq.~(\ref{eq:white_noise_sensitivity}) with $K=\sqrt{2}$.
    
\begin{figure*}
         \begin{center}
             \includegraphics[width=17cm]{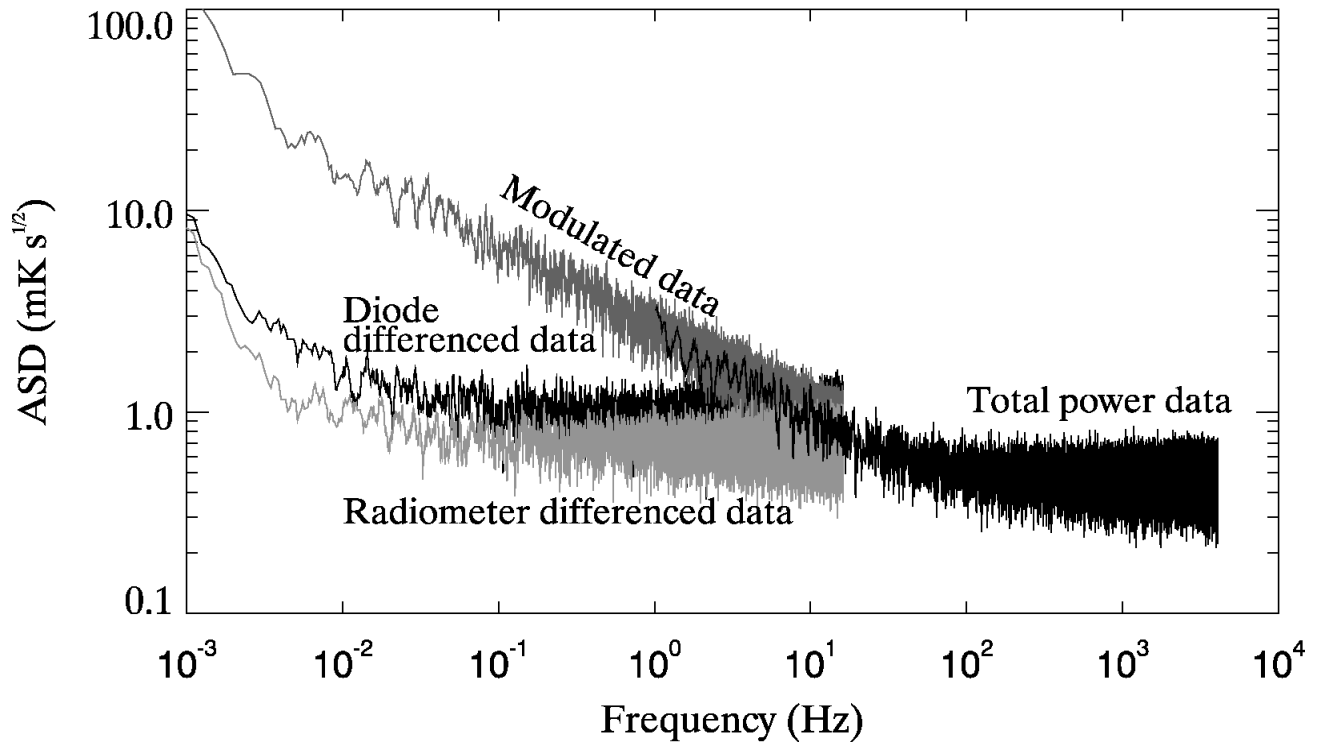}
         \end{center}
        \caption{
            Amplitude spectral densities of unswitched and differenced data streams. The pseudo-correlation differential
            design reduces the $1/f$ knee frequency by three orders of magnitude. The white noise level scales almost
            perfectly with $K$.
        }
        \label{fig:example_spectral_densities}
    \end{figure*}

    Fig.~\ref{fig:example_spectral_densities} shows the effectiveness of the LFI pseudo-correlation design
    \citep[see][]{2009_LFI_cal_R2}. The $1/f$ knee frequency is reduced, after differencing, by more than three orders
    of magnitude, and the white noise sensitivity scales almost perfectly with the three values of the constant $K$. 
   The following terminology is used in the figure:
    \begin{itemize}
     \item \textit{Total power data}: datastreams acquired without operating the phase switch;
     \item \textit{Modulated data}: datastreams acquired in nominal, switching conditions before taking the
difference in Eq.~\ref{eq:ideal_power_output};
     \item \textit{Diode differenced data}: differenced datastreams for each diode;
     \item \textit{Radiometer differenced data}: datastreams obtained from a weighted average of
the two diode differenced datastreams for each radiometer (see Eq.~\ref{eq:weighted_diode_averaging}).
    \end{itemize}
    
    \subsubsection{Overview of main noise parameters}
    \label{sec:overview_main_noise_parameters}
    
        If we consider a typical differenced data noise power spectrum, $P(f)$, we can identify three main
characterisics:
    
        \begin{enumerate}
            \item the white noise plateau, where $P(f)\sim \sigma^2$. The white noise sensitivity is given by $\sigma$
(in units of K\,s$^{1/2}$), and the noise effective bandwidth by:
            \begin{equation}
                \beta = {(K V_{\rm DC} / \sigma_{\rm V})^2 \over  \left[ 1 + b\, G_0(T_{\rm sky}+T_{\rm
noise})\right]^{2}},
                \label{eq:noise_eff_bw}
            \end{equation}
            where $V_{\rm DC}$ is the voltage DC level, $\sigma_{\rm V}$ the uncalibrated white noise sensitivity and
the term in square brackets represents the effect of compressed voltage output (see
Appendix~\ref{app:noise_eff_bw_calculation});
    
            \item the $1/f$ noise tail, characterised by a power spectrum $P(f)\sim \sigma^2 (f/f_k)^{-\alpha}$
described by two parameters: the knee frequency, $f_k$, defined as the frequency where the $1/f$ and white noise
contribute equally, and the slope $\alpha$;
    
            \item spurious frequency spikes. These are a common-mode additive effect caused by interference between
scientific and housekeeping data in the analog circuits of the data acquisition electronics box (see
Sect.~\ref{sec:spikes}).
        \end{enumerate}

    \subsubsection{Test experimental conditions}
    \label{sec:test_experimental_conditions}
    
        The test used to determine instrument noise was a long duration (2 days) acquisition during which the instrument
ran undisturbed in its nominal mode. Target temperatures were set at $T_{\rm sky} = 19$\,K and $T_{\rm ref} =
22$\,\hbox{K}. The front-end unit was at 26\,K, maintained stable to $\pm 10$~mK.
    
\begin{figure}[h!]
             \begin{center} 
                 \includegraphics[width = 8.8 cm]{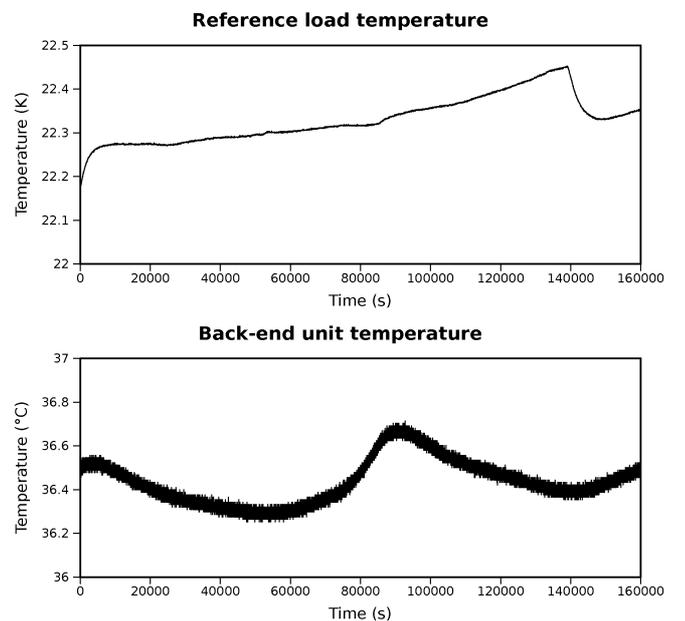}
             \end{center} 
            \caption{
                Thermal instabilities during the long duration acquisition. {\it Top:} drift in the reference load
temperature caused by leakage in the gas cap thermal switch. The drop towards the end of the test coincides with refill
of the thermal switch. {\it Bottom:} 24-hr back-end temperature fluctuation. 
            }
            \label{fig:thermal_instabilities_st1_0002}
        \end{figure}

        The most relevant instabilities were a 0.5\,K peak-to-peak 24-hour fluctuation in the back-end temperature and a
200\,mK drift in the reference load temperature caused by a leakage in the gas gap thermal switch that was refilled
during the last part of the acquisition (see Fig.~\ref{fig:thermal_instabilities_st1_0002}).
    
\begin{figure*}[h!]
             \includegraphics[width=17cm]{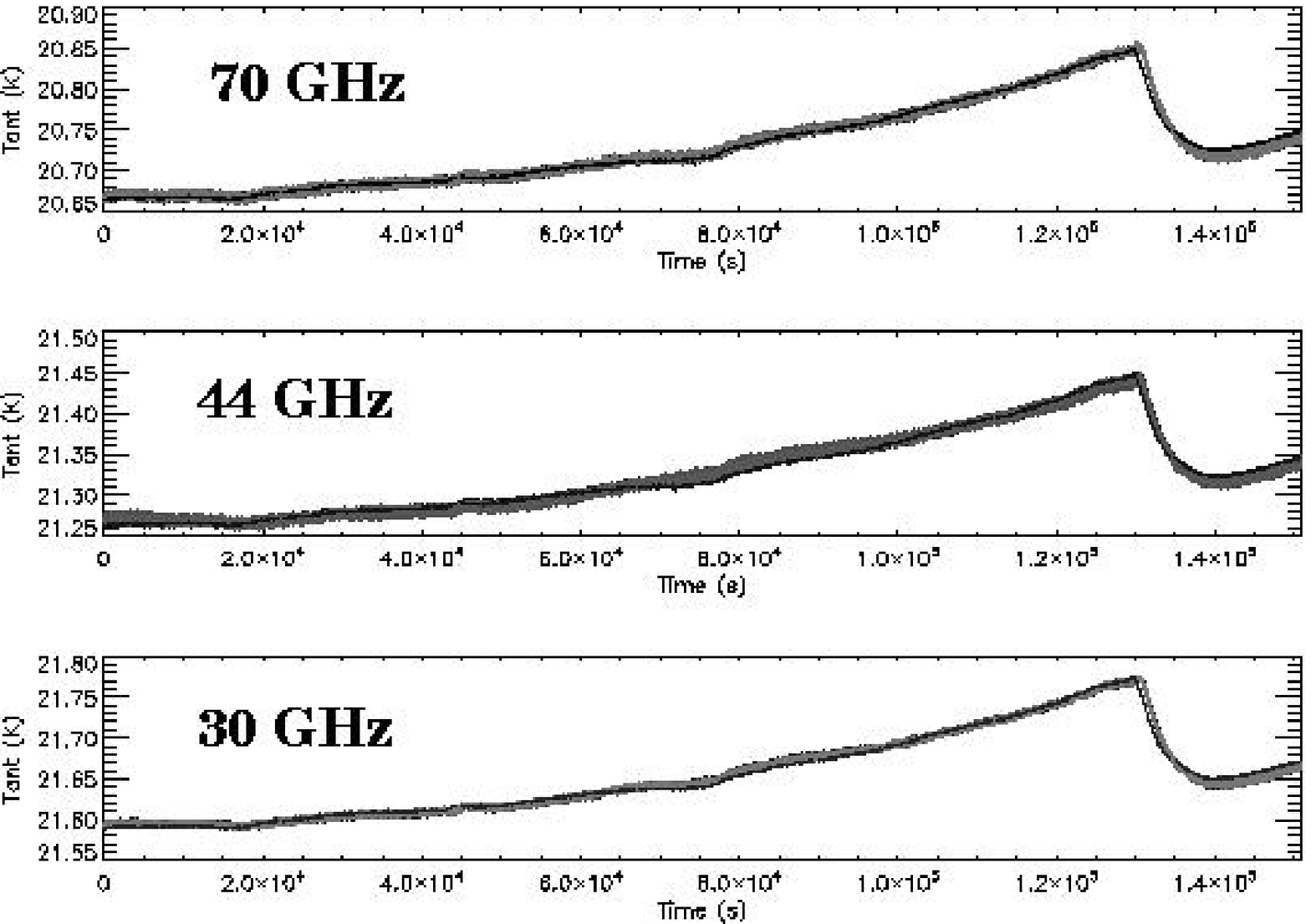}
            \caption{
                Calibrated differential radiometric outputs (downsampled to 1\,Hz) for all LFI detectors during the long
duration test. Temperature sensor data in antenna temperature units are superimposed (thin black line) on the calibrated
radiometric data.
            }
            \label{fig:differential_output}
        \end{figure*}

        The effect of the reference load temperature variation was clearly identified in the differential radiometric
output (see Fig.~\ref{fig:differential_output}) and  removed from the radiometer data before differencing.  The effect
of the back-end temperature was removed by correlating the radiometric output with temperature sensor measurements.
    
    \subsubsection{White noise sensitivity and noise effective bandwidth}
    \label{sec:white_noise_eff_bw}
    
        There are four sources of the white noise that determines the final sensitivity: (i) the input sky signal; (ii)
the RF part of the receiver (active components and resistive losses); (iii) the back-end electronics after the detector
diode\footnote{The additional noise introduced by the analog electronics is generally negligible compared to the
intrisic noise of the receiver, and its impact was further mitigated by the variable gain stage after the diode.}; and
(iv) signal quantisation performed in the digital processing unit. 
    
        Signal quantisation can increase significantly the noise level if $\sigma / q \lesssim 1$, where $q$ represents
the quantisation step and $\sigma$ the noise level before quantisation.  Previous optimisation studies \citep{maris03}
have shown that a quantisation ratio $\sigma / q \sim 2$ is enough to satisfy telemetry requirements without
significantly increasing the noise level. This has been verified during calibration tests using the so-called
``calibration channel'', i.e., a data channel containing about 15 minutes per day of unquantised data from each
detector. The use of the calibration channel allowed a comparison between the white noise level before and after
quantisation and compression for each detector. Table~\ref{tab:white_noise_before_after_compression} summarises these
results and shows that digital quantisation caused an increase in the signal white noise less than 1\%.
        
\begin{figure}[h!]
             \begin{center}
                 \includegraphics[width=9cm]{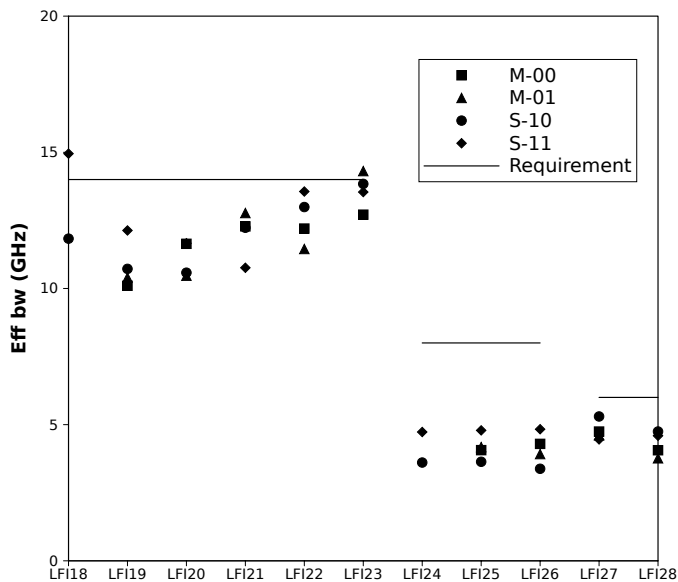}
             \end{center}
    
            \caption{
                Noise effective bandwidths calculated during RAA measurements. The three lines indicate the 70~GHz,
44~GHz and 30~GHz requirements.
            }
            \label{fig:noise_effective_bandwidths}
        \end{figure}

        The white noise effective bandwidth calculated according to Eq.~(\ref{eq:noise_eff_bw}) is reported in
Fig.~\ref{fig:noise_effective_bandwidths}. Our results indicate that the noise effective bandwidth is smaller than the
requirement by 20\%, 50\%, and 10\% at 30, 44, and 70\,GHz, respectively. Non-idealities in the in-band response
(ripples) causing bandwidth narrowing are discussed in \citet{2009_LFI_cal_R3}.
    
        It is useful to extrapolate these results to the expected in-flight sensitivity with the instrument at the
nominal temperature of 20\,K and observing a sky signal of $\sim 2.73$\,K in thermodynamic temperature.  This estimate
has been performed in two different ways.  The first uses measured noise effective bandwidths and noise temperatures in
the radiometer equation, Eq.~(\ref{eq:white_noise_sensitivity}). The second starts from measured uncalibrated noise,
which is then calibrated in temperature units, corrected for the different focal plane temperature in test conditions,
and extrapolated to $\sim 2.73$~K input using the radiometeric response equation, Eq~(\ref{eq:nonlinear_response}). The
details of the extrapolation are given in Appendix~\ref{app:white_noise_calibration_extrapolation}.
    
        Table~\ref{tab:noise_per_radiometer} gives the sensitivity per radiometer estimated according to the two
procedures. The sensitivity per radiometer has been obtained using a weighted noise average from the two detectors of
each radiometer (see Appendix~\ref{app:weighted_noise_average}). Because radiometers \texttt{LFI18M-0} and
\texttt{LFI24M-0} were not working during the tests, we have estimated the sensitivity per frequency channel considering
the white noise sensitivity of \texttt{LFI24M-0} (that was later repaired) measured during receiver-level tests while
for \texttt{LFI18M-0} (that was later replaced with a spare unit) we have assumed the same sensitivity of
\texttt{LFI18S-1}. Further details about white noise sensitivity of individual detectors are reported in
\citet{2009_LFI_cal_R2}.
    
        \begin{table}
            \caption{
                Calibrated white noise sensitivities in $\mu$K\,s$^{1/2}$ per radiometer extrapolated at CMB input using
the two methods outlined above and detailed in Appendix~\ref{app:white_noise_calibration_extrapolation}.
            }
            \label{tab:noise_per_radiometer}
            \begin{center}
                \begin{tabular}{l c c}
                    \hline
                    \hline
                    \multicolumn{3}{c}{\textbf{From uncalib. noise}}\\
                    &M-0 & S-1 \\ 
                    \hline
                    \textbf{70~GHz}  &     &  \\       
                    LFI18   &468    &468   \\
                    LFI19   &546    &522   \\
                    LFI20   &574    &593   \\
                    LFI21   &424    &530   \\
                    LFI22   &454    &463   \\
                    LFI23   &502    &635   \\
                    \hline
                    \textbf{44~GHz}  &     &  \\       
                    LFI24   &372    &447    \\
                    LFI25   &501    &492    \\
                    LFI26   &398    &392    \\
                    \hline
                    \textbf{30~GHz}  &     &  \\       
                    LFI27   &241     &288   \\
                    LFI28   &315     &251   \\        
                    \hline
                \end{tabular}
                \begin{tabular}{l c c }
                    \hline
                    \hline
                    \multicolumn{3}{c}{\textbf{From radiom. equation}}\\
                    &M-0 & S-1 \\ 
                    \hline
                    \textbf{70~GHz}  &   &  \\       
                    LFI18   &450    &450\\
                    LFI19   &482    &466\\
                    LFI20   &498    &511\\
                    LFI21   &381    &496\\
                    LFI22   &428    &410\\
                    LFI23   &453    &419\\
                    \hline
                    \textbf{44~GHz}  &   &  \\       
                    LFI24   &404    &407\\
                    LFI25   &451    &462\\
                    LFI26   &455    &428\\
                    \hline
                    \textbf{30~GHz}  &    &  \\       
                    LFI27   &311    &320\\
                    LFI28   &305    &268\\        
                    \hline
                \end{tabular}
            \end{center}
        \end{table}

        The sensitivity per frequency channel estimated according the two procedures and compared with the LFI
requirement is shown in Table~\ref{tab:calibrated_white_noise_results}.

        \begin{table}[h!]
            \caption{
                Calibrated white noise sensitivities in $\mu$K\,s$^{1/2}$ per frequency channel extrapolated at CMB
input using the two methods outlined above and detailed in Appendix~\ref{app:white_noise_calibration_extrapolation}. The
third column reports the LFI requirement.
            }
            \label{tab:calibrated_white_noise_results}
            \begin{center}
                \begin{tabular}{l p{1cm} p{1cm} p{1cm}}
                \hline
                \hline
                    &\textbf{Meas. noise}       &\textbf{Rad. eq.}      &\textbf{Req.}\\
                \hline
                70 GHz  &146    &130    &105    \\
                44 GHz  &174    &177    &113    \\
                30 GHz  &135    &149    &116    \\
                \hline
                \end{tabular}
            \end{center}
        \end{table}

    \subsubsection{1/$f$ noise parameters}
    \label{sec:one_over_f}
    
        The $1/f$ noise properties of the LFI differenced data have been better determined during instrument-level than
        receiver-level tests for two reasons: (i) the test performed in this phase has been the longest in all the test
        campaign, and (ii) because of the better temperature stability, especially compared to the 70~GHz receivers
        cryofacility \citep{2009_LFI_cal_M4}.
    
best fit with two 1/$f$ components. Being a second-order correction we limit, in this paper, to results obtained with
the simple model that
        Results, summarised in Table~\ref{tab:knee_fequency_slope}, show very good $1/f$ noise stability of the LFI
receivers, almost all with a knee frequency well below the required 50\,mHz.
    
        \begin{table}
            \caption{
                Summary of knee frequency and slope for all LFI detectors.
            }
            \label{tab:knee_fequency_slope} 
            \begin{center}
                \begin{tabular}{l c c c c}
                \hline
                \hline
                        &\multicolumn{4}{c}{$f_{\rm knee}$ (mHz)}\\           
                        &\textbf{M-00} 
                        &\textbf{M-01} 
                        &\textbf{S-10} 
                        &\textbf{S-11} \\
                \hline
                \textbf{70~GHz}  &   &   &   &  \\       
                        LFI18   &$\ldots$ &$\ldots$ &61 &59\\
                        LFI19   &25 &32 &27 &37\\
                        LFI20   &21 &19 &23 &28\\
                        LFI21   &28 &30 &41 &38\\
                        LFI22   &46 &39 &41 &76\\
                        LFI23   &30 &31 &58 &75\\
                \hline      
                \textbf{44~GHz}  &   &   &   &  \\       
                        LFI24   &$\ldots$ &$\ldots$ &39 &46\\
                        LFI25   &31 &31 &21 &30\\
                        LFI26   &61 &61 &61 &14\\
                \hline      
                \textbf{30~GHz}  &   &   &   &  \\       
                        LFI27   &30 &30 &27 &26\\
                        LFI28   &37 &31 &37 &39\\
                \hline
                \end{tabular}
                
            \begin{tabular}{l c c c c}
                \hline
                        &\multicolumn{4}{c}{slope}\\    
                        &\textbf{M-00} 
                        &\textbf{M-01} 
                        &\textbf{S-10} 
                        &\textbf{S-11} \\
                \hline
                \textbf{70~GHz}  &   &   &   &  \\       
                    LFI18       &$\ldots$       &$\ldots$       &$-$1.12        &$-$1.12\\
                    LFI19       &$-$1.27        &$-$1.22        &$-$1.11        &$-$1.02\\
                    LFI20       &$-$1.47        &$-$1.64        &$-$1.27        &$-$1.24\\
                    LFI21       &$-$1.48        &$-$1.61        &$-$1.15        &$-$1.17\\
                    LFI22       &$-$1.18        &$-$1.26        &$-$1.19        &$-$1.01\\
                    LFI23       &$-$1.11        &$-$1.19        &$-$1.15        &$-$1.12\\
                \hline
                \textbf{44~GHz}  &   &   &   &  \\       
                    LFI24       &$\ldots$       &$\ldots$       &$-$1.06        &$-$1.11\\
                    LFI25       &$-$1.07        &$-$1.03        &$-$1.10        &$-$1.00\\
                    LFI26       &$-$1.01        &$-$1.01        &$-$1.05        &$-$1.55\\
                \hline
                \textbf{30~GHz}  &   &   &   &  \\       
                    LFI27       &$-$1.06        &$-$1.13        &$-$1.25        &$-$1.13\\
                    LFI28       &$-$0.94        &$-$0.93        &$-$1.07        &$-$1.06\\
                \hline
                \end{tabular}
            \end{center}
    
        \end{table}

    \subsubsection{Spurious frequency spikes}
    \label{sec:spikes}
    
        During the FM test campaign we found unwanted frequency spikes in the radiometeric data at frequencies of the
order of few hertz. The source of the problem was recognised to be in the backend data acquisition electronics box,
where  unexpected crosstalk between the circuits handling housekeeping and radiometric data affected the radiometer
voltage output downstream of the detector diode.
    
\begin{figure}[]
            \begin{center}
                 (a)\includegraphics[width=8cm]{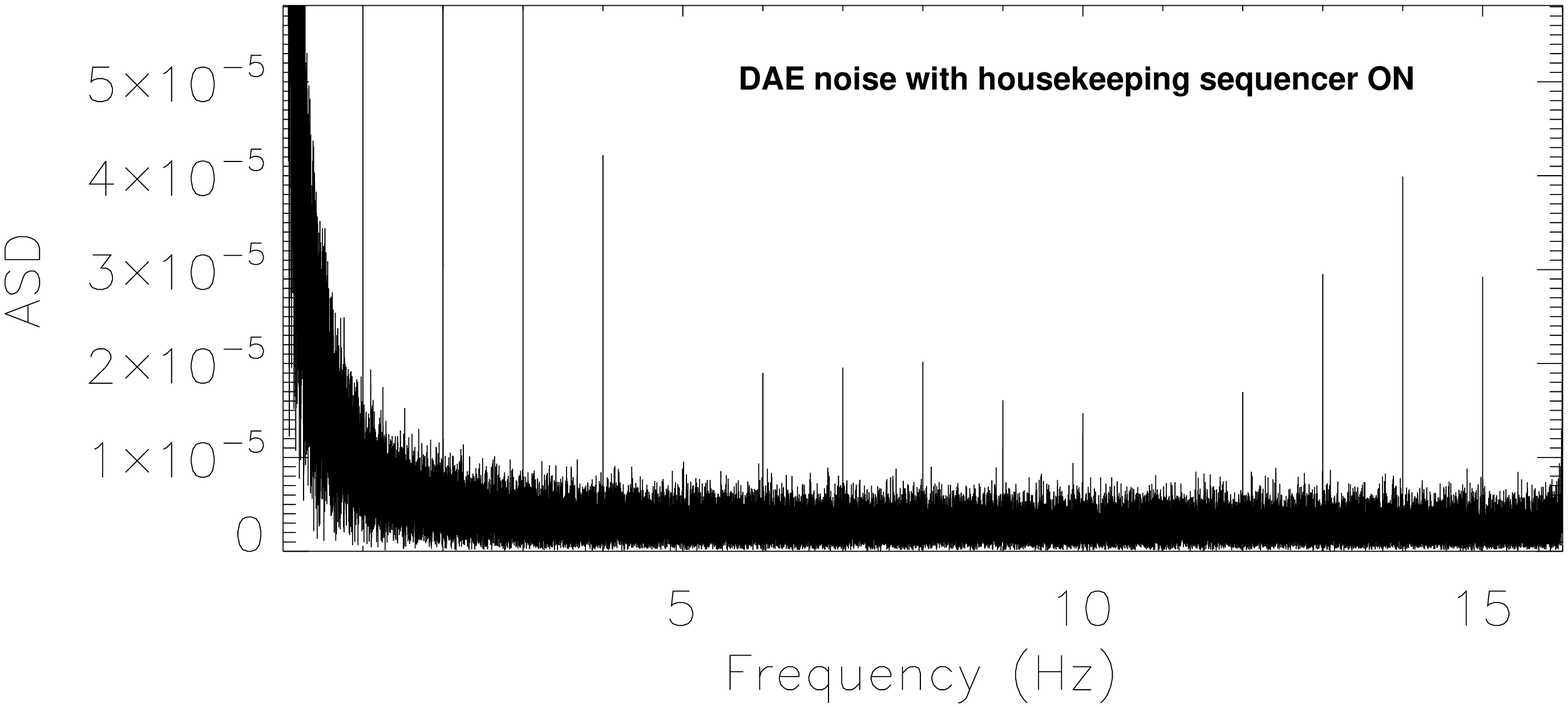} \\
                 \vspace{0.4cm}
                 (b)\includegraphics[width=8cm]{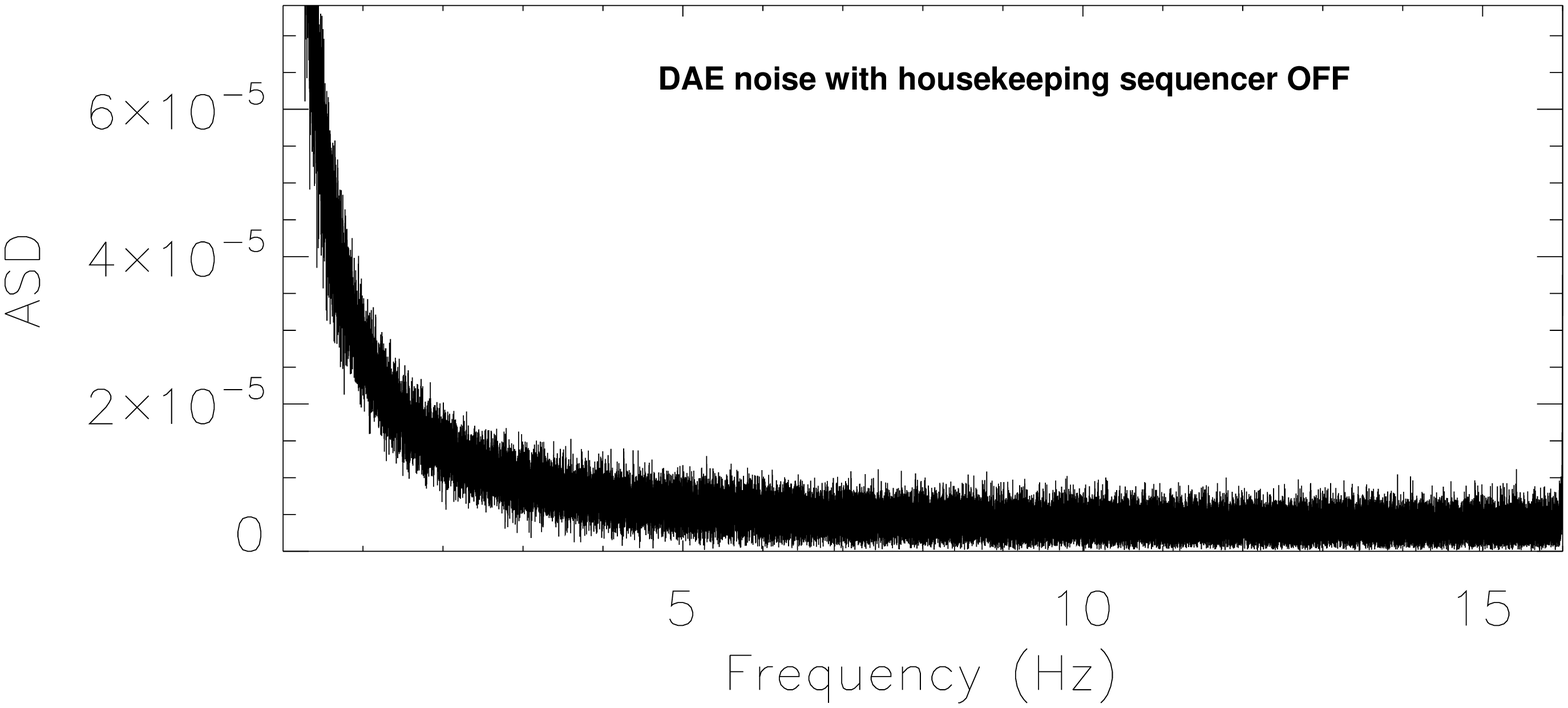} \\
                 \vspace{0.4cm}
                    \mbox{}\\
                 (c)\includegraphics[width=8cm]{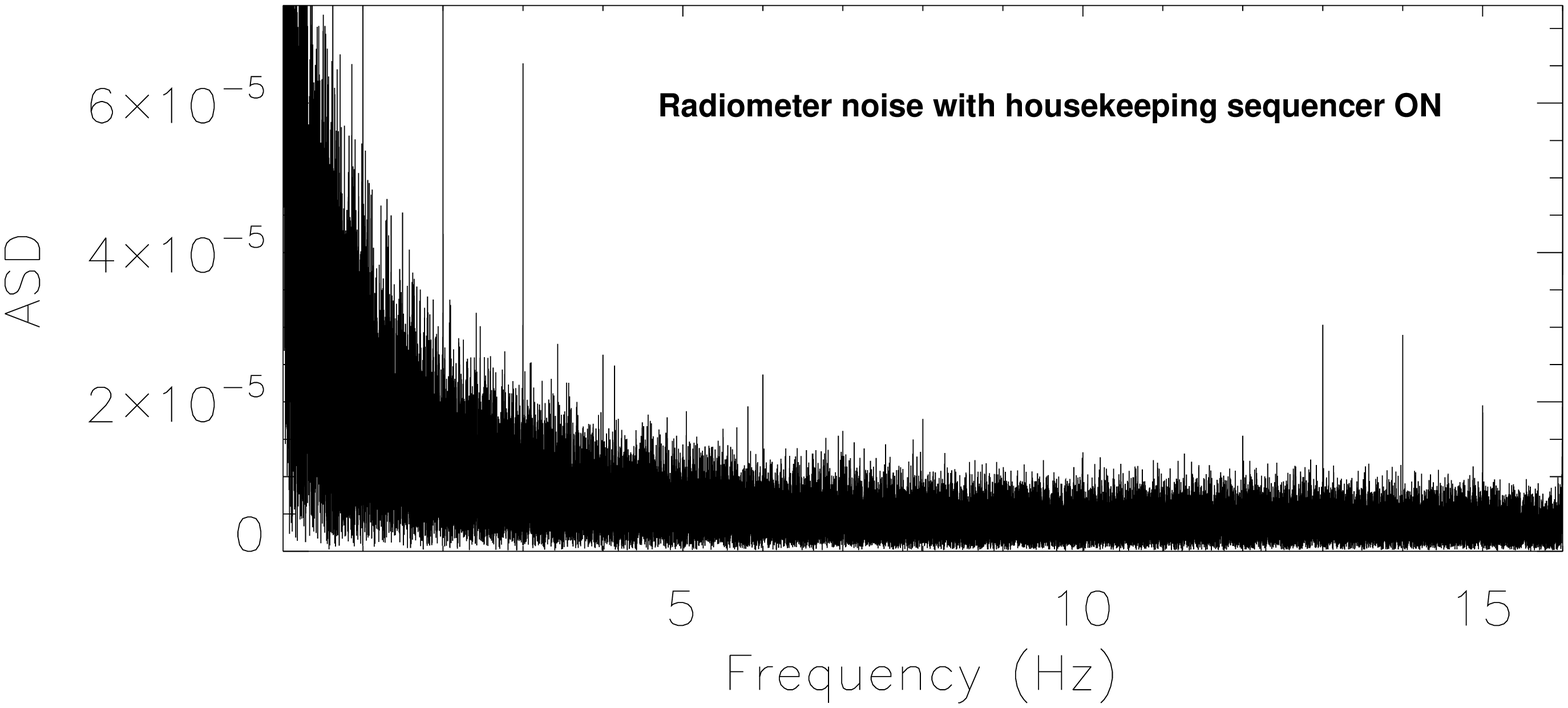} \\
                 \vspace{0.4cm}
                 (d)\includegraphics[width=8cm]{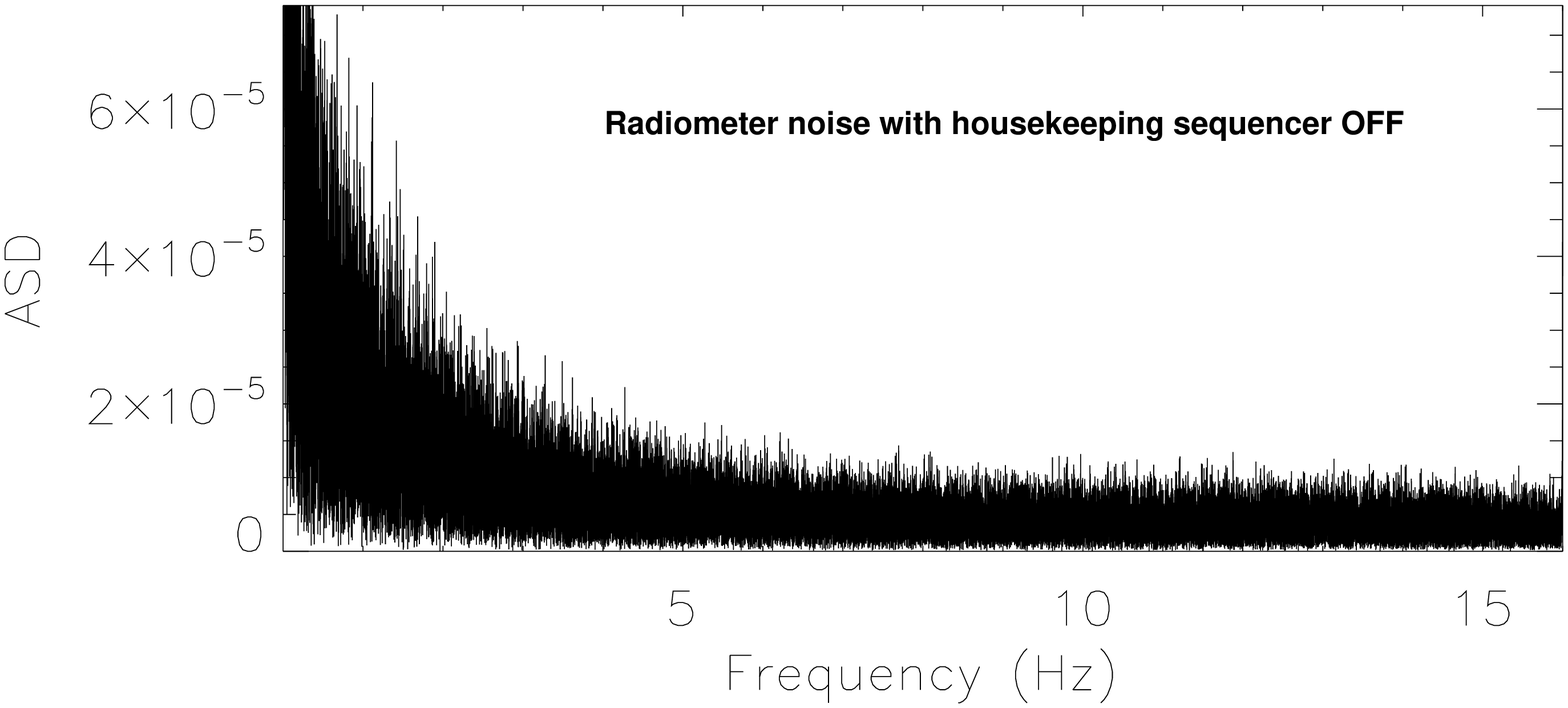} 
            \end{center}
    
            \caption{
                DAE-only and radiometer noise amplitude density spectra in V$/\sqrt{\rm Hz}$ (from \texttt{LFI18S-10}
unswitched data) with and without activation of the housekeeping acquisition. These data show clearly that the source of
the disturbance is in the data acquisition electronics box and is correlated with the status of the housekeeping data
acquisition.
            }
            \label{fig:frequency_spikes_hk_on_off}
        \end{figure}

        This is shown clearly in Fig.~\ref{fig:frequency_spikes_hk_on_off}, which shows spectra of unswitched data
acquired from the 70\,GHz detector \texttt{LFI18S-10} with the housekeeping data acquisition activated and deactivated. 
    
        Because the disturbance is added to receiver signal at the end of the radiometric chain it acts as a common
mode effect on both the sky and reference load data so that its effect in differenced data is reduced by several orders
of magnitude bringing it well below the radiometer noise level.
    
        Further analysis of these spikes has shown that the disturbance is synchronized in time. By binning the data
synchronously we obtain a template of the disturbance, which allows its removal in time-domain \citep{2009_LFI_cal_R2}.
The feasibility of this approach has been proven with data acquired during the full satellite test campaign in
Liege, Belgium
during July and August, 2008.
 
        Therefore, because the only way to eliminate the disturbance \textit{in hardware} would be to operate
the instrument without any housekeeping information, our baseline approach is that, if necessary, the
residual effect will be removed from the data in time domain after measuring the disturbance shape from the flight
data.


\subsection{Radiometric suceptibility to front-end temperature instabilities}
\label{sec:susceptibility}

Thermal fluctuations in the receivers result in gain changes in the amplifiers and noise changes in the (slightly
emissive) passive components (horns, OMTs, waveguides).  These changes mimic the effect of changes in sky emission,
expecially at fluctuation frequencies near the satellite spin frequency.  The most important source of temperature
fluctuations for LFI is the sorption cooler \citep{Bhandari04:planck_sorption_cooler, wade00}.  

    For small temperature fluctuations in the focal plane the radiometric response is linear \citep{seiffert02,
2009_LFI_cal_R6}, so the spurious antenna temperature fluctuation in the differential receiver output can be written as:
   
    \begin{equation}
            \delta T_{\rm out} = f_{\rm trans} \delta T_{\rm phys}.
            \label{eq:susceptibility_general_equation} 
    \end{equation}
    Transfer function $f_{\rm trans}$ can be estimated analytically from the differential power output given in
Eq.~(\ref{eq:ideal_power_output}):
   
    \begin{equation}
            f_{\rm trans} = \frac{\partial p_{\rm out}}{\partial T_{\rm phys}} 
            \left(\frac{\partial p_{\rm out}}{\partial T_{\rm sky} }\right)^{-1}.
        \label{eq:susceptibility_transfer_function}
    \end{equation}
   
    The analytical form of $f_{\rm trans}$ \citep[discussed in detail in][]{2009_LFI_cal_R6} depends primarily on the
front-end amplifier susceptibility parameters, $\partial G / \partial T_{\rm phys}$ and $\partial T_{\rm noise} /
\partial T_{\rm phys}$ as well as on other instrument and boundary condition parameters such as the insertion loss of
passive components and the sky input temperature.
    
    If we consider the systematic error budget in \citet{2009_LFI_cal_M2}, it is possible to derive a requirement for
the radiometric transfer function, $f_{\rm trans} \lesssim 0.1$, in order to maintain the final peak-to-peak error per
pixel $\lesssim 1 \mu$K (see Appendix~\ref{app:susceptibility_scientific_requirement}). During instrument-level
calibration activities dedicated tests were run to estimate $f_{\rm trans}$ and compare it with theoretical estimates
and similar tests performed on individual receivers.
   
    \subsubsection{Experimental setup}
    \label{sec:susceptibility_measurements}
   
\begin{figure}[h!]
             \includegraphics[width=8.5cm]{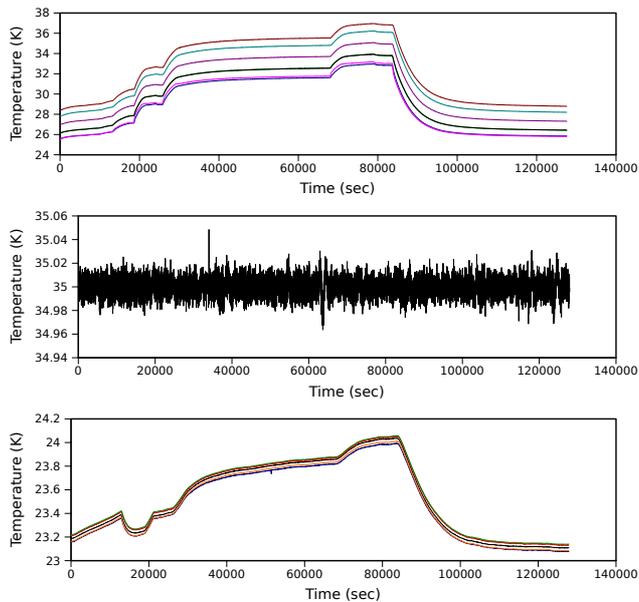}
            \caption{
                Behaviour of focal plane (top), sky load (middle) and reference load (bottom) temperatures during the
thermal susceptibility tests.
            }
            \label{fig:thf_temperatures}
        \end{figure}

        During this test the focal plane temperature was varied in steps between between 27 and 34\,K. The sky and
reference load temperatures were $T_{\rm sky}=35\pm0.01$\,K and $T_{\rm ref}=23.7\pm0.01$\,\hbox{K}. The reference load
temperature showed a non-negligible coupling with the focal plane temperature (as shown in
Fig~\ref{fig:thf_temperatures}) so that the effect of this variation had to be removed from the data before calculating
the thermal transfer function.

        Although the test lasted more than 24\,hours, it was difficult to reach a clean steady state plateau after each
step because of the large thermal mass of the instrument.  Furthermore, for some detectors the bias tuning was not yet
optimised, so that only data from a subset of detectors could be compared with similar measurements performed at
receiver-level.

\begin{figure}[h!]
             \begin{center}
                    \includegraphics[width=4.2cm]{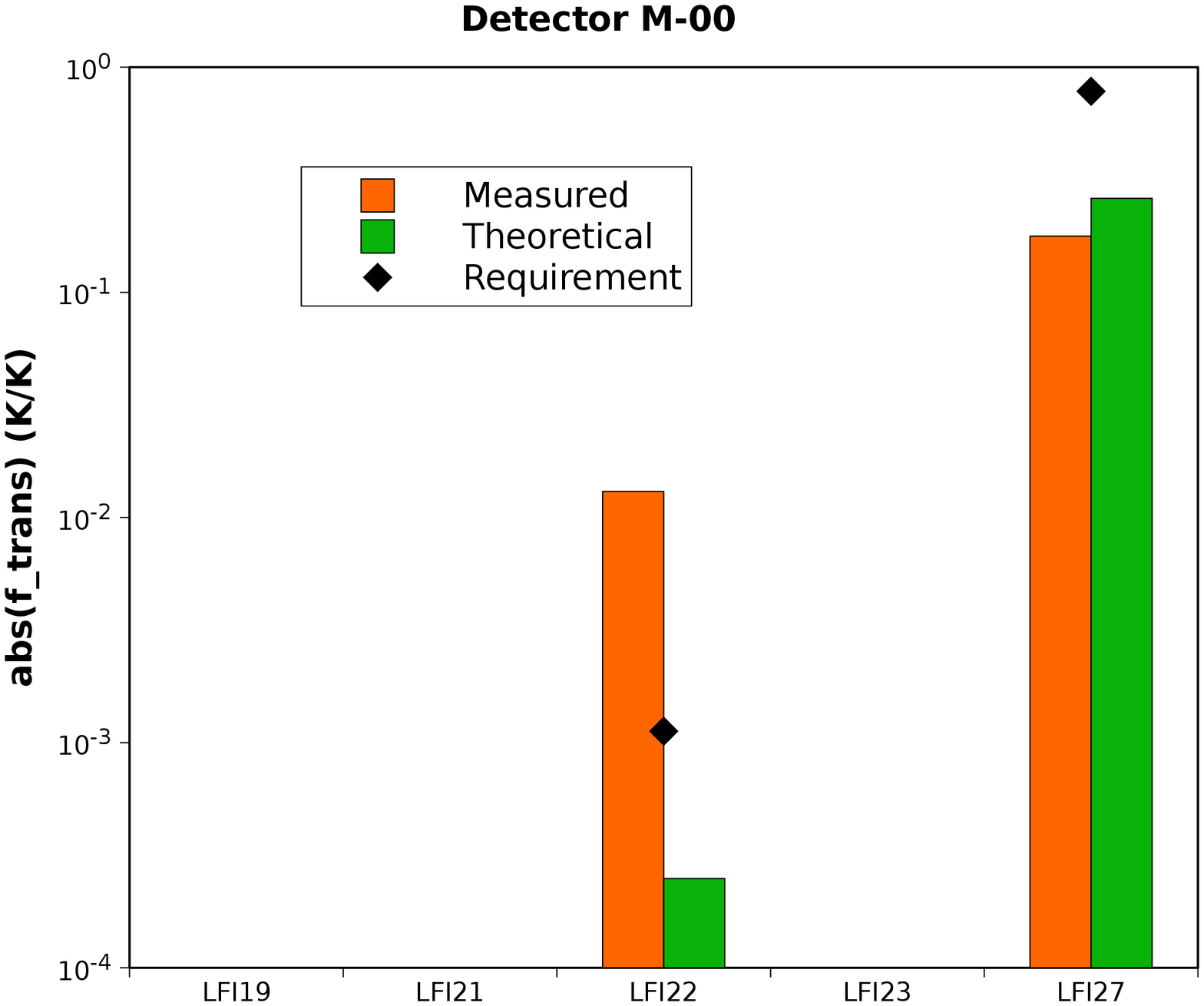}
                    \includegraphics[width=4.2cm]{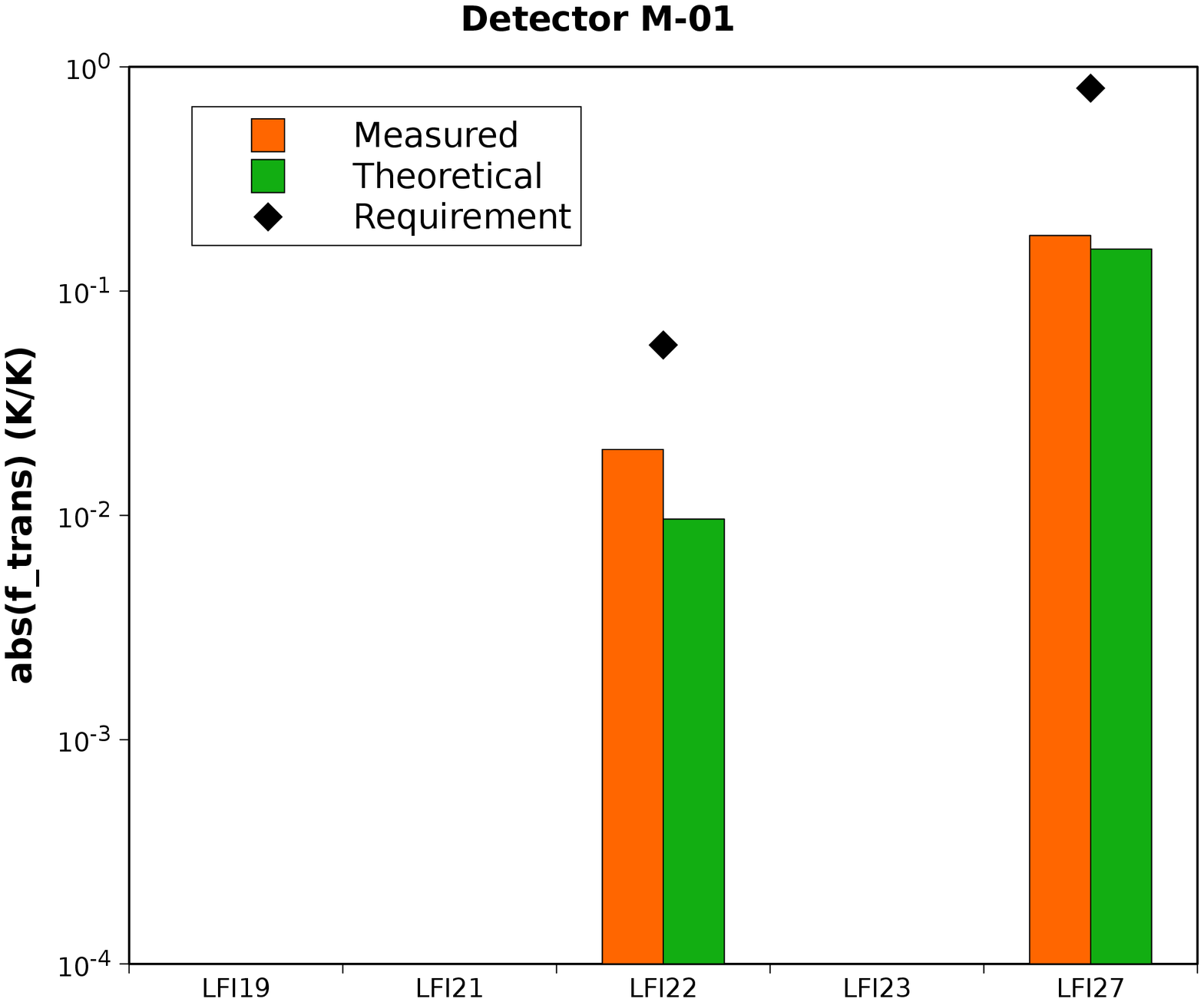}\\
                 \vspace{0.5cm}
                    \includegraphics[width=4.2cm]{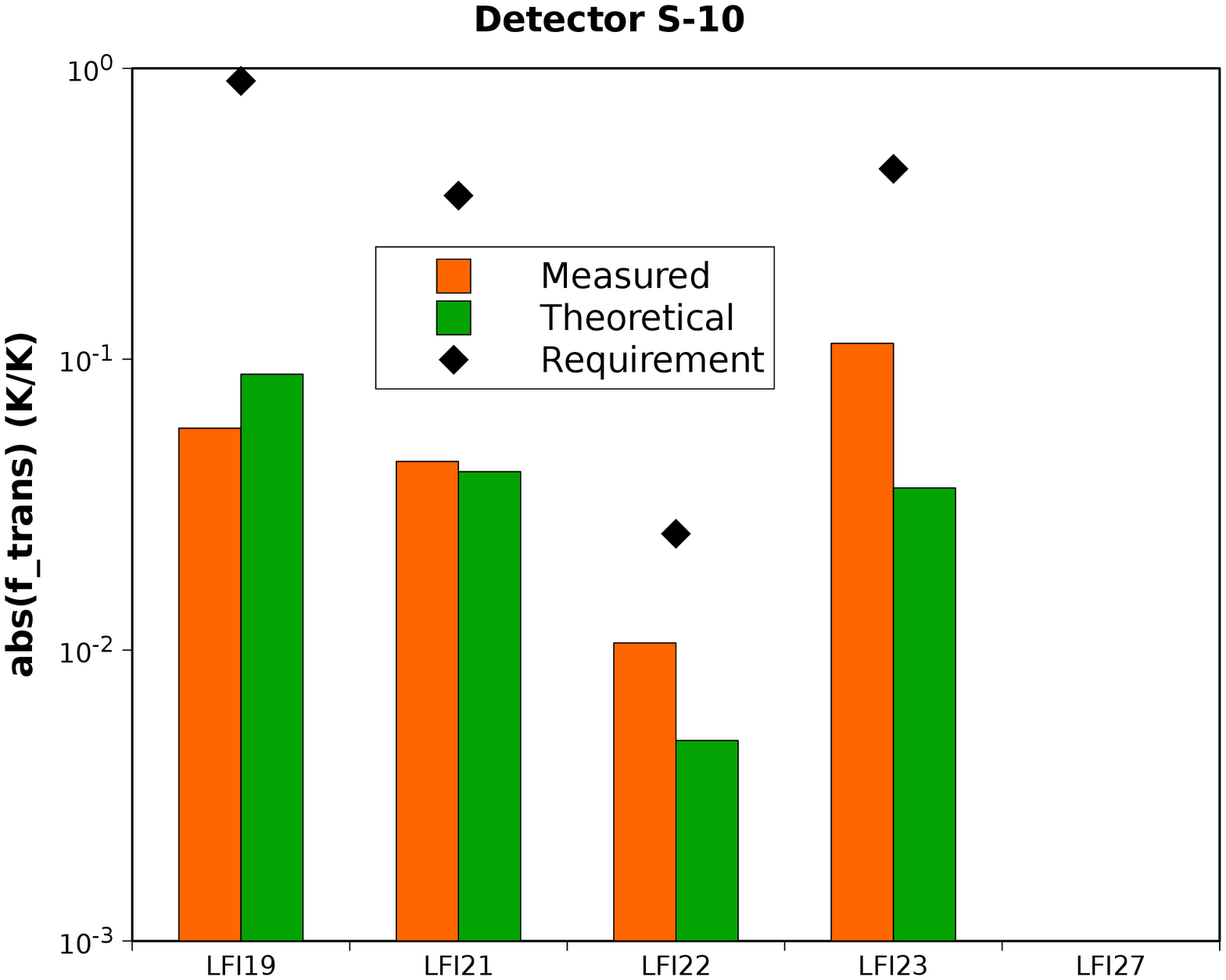}
                    \includegraphics[width=4.2cm]{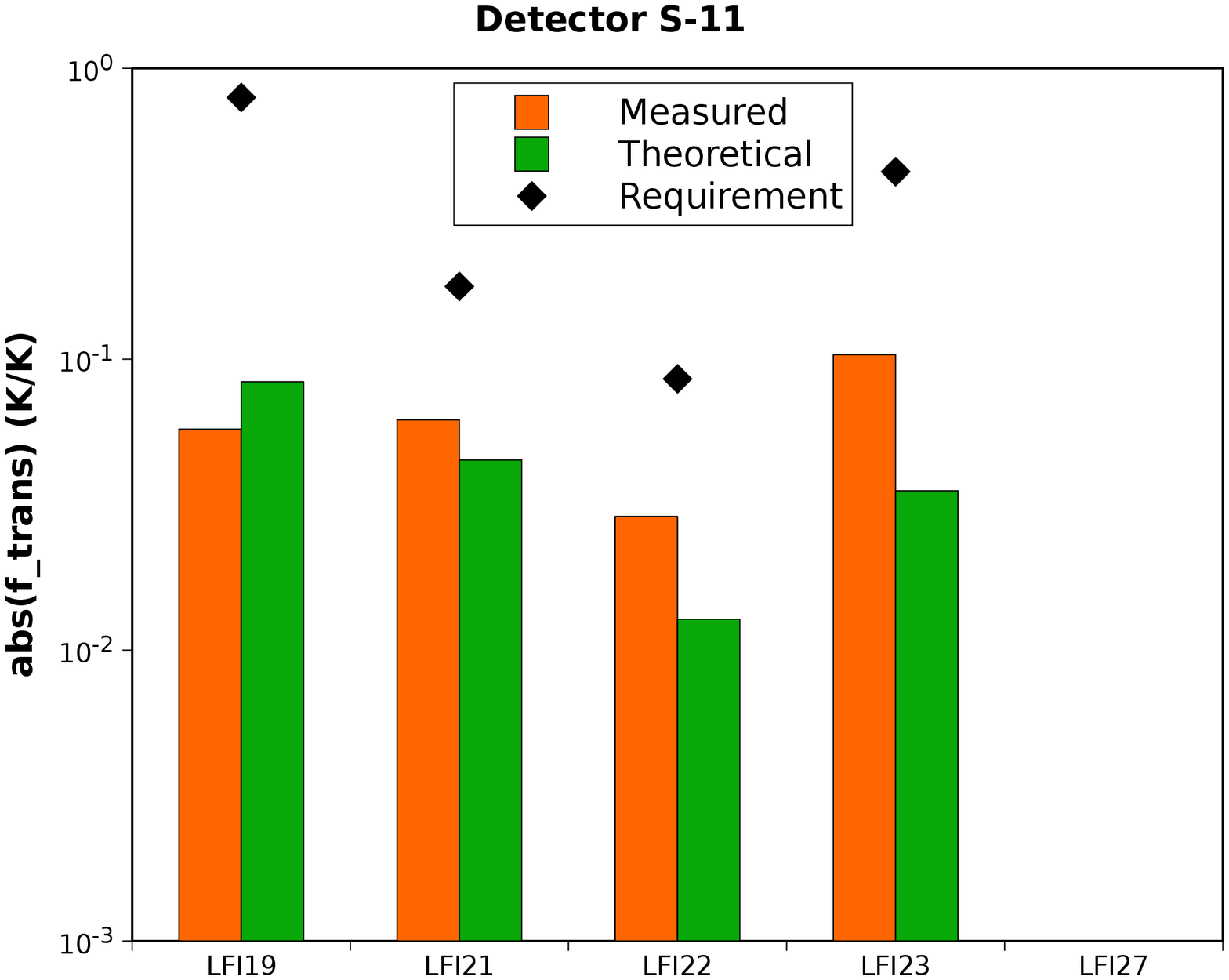}
                \end{center}
                \caption{
                Measured and predicted radiometric thermal transfer functions, with the scientific requirement rescaled
at the experimental conditions of the test. The comparison is possible only for the subset of radiometers that was tuned
at the time of this test.
            }
                \label{fig:thf_results}
        \end{figure}

        In Fig.~\ref{fig:thf_results} we summarise our results by comparing predicted and measured transfer functions
for the tested detectors. Predicted transfer functions have been calculated using the list of parameters provided in
Appendix~\ref{app:susceptibility parameters}, derived from receiver-level tests. In the same figure we also plot the
thermal susceptibility requirement rescaled at the experimental test conditions with a scale factor given by the ratio 
        
        \begin{equation}
        f_{\rm trans}({\rm ground}) / f_{\rm trans}({\rm flight}),
        \end{equation}
        where $f_{\rm trans}$ has been calculated using Eq.~(\ref{eq:susceptibility_general_equation}) in ground and
flight conditions from sky, reference-load and focal plane temperatures.

        Fig.~\ref{fig:thf_results} shows that transfer functions measured during instrument-level tests are compliant
with scientific requirements and reflect theoretical predictions, with the exception of \texttt{LFI22} and
\texttt{LFI23}, which were more susceptible to front-end temperature fluctuations than expected. In general, results
from the instrument test campaign confirm design expectations, and suggest that the level of temperature instabilities
in the focal plane will not represent a significant source of systematic errors in the final scientific products. This
has been further verified during satellite thermal-vacuum tests conducted with the flight model sorption cooler (see
Sect.~\ref{sec:csl_thermal_suceptibility}).

\section{Comparison with satellite level test results}
\label{sec:satellite_results}

The final cryogenic ground test campaign was conducted at the Centre Spatiale de Li\`ege (CSL) with the LFI and the HFI
integrated on board Planck. To reproduce flight temperature conditions, the satellite was enclosed in an outer
cryochamber cooled to liquid nitrogen temperatures, and surrounded an inner thermal shield at $\sim$20\,\hbox{K}. An
ECCOSORB load cooled to 4.5\,K was placed between the secondary mirror and the feed horns to simulate the cold sky. For
the first time, the LFI focal plane was cooled to 20\,K by the sorption cooler, and the reference loads were cooled to
$\sim$4\,K by the 4\,K cooler.

During the CSL tests we verified instrument functionality, tuned front-end biases and back-end electronics, and assessed
scientific performance in the closest conditions to flight attainable on the ground. Front-end bias tuning made use of
the ability of the 4\,K cooler system to provide several different stable temperatures to the reference loads in the
range 24\,K down to the nominal 4\,\hbox{K} \citep{2009_LFI_cal_R7}.

A detailed description of satellite-level tests is outside the scope of this paper; here we will focus on the comparison
of the main performance parameters measured during instrument and satellite tests, and show that despite differences in
test conditions the overall behaviour was reproduced.

\subsection{White noise sensitivity}
\label{sec:csl_white_noise}

    Calibrated white noise sensitivities were determined during satellite-level tests by exploiting a $\sim 80$\,mK
variation of the sky load temperature caused by the periodic helium refills of the chamber.  This variation allowed us
to estimate the photometric calibration constant by correlating the differenced voltage datastream $\delta V(t)$ from
each detector with the sky load temperature $T_{\rm sky}^{\rm ant}(t)$ (in antenna temperature units). 

    To extrapolate the calibrated sensitivity from the 4.5\,K input temperature in the test to flight conditions we
calculated the ratio:
    
    \begin{equation}
        \frac{\Delta T_{\rm rms}(T_{\rm sky}^{\rm flight})}{\Delta T_{\rm rms}(T_{\rm sky}^{\rm CSL})} = \frac{(T_{\rm
sky}^{\rm flight} + T_{\rm noise})}{(T_{\rm sky}^{\rm CSL} + T_{\rm noise})}
        \label{eq:ratio_flight_csl}
    \end{equation}
    using the noise temperature found from the non-linear model fit from the receiver-level test campaign
\citep{2009_LFI_cal_M4}. This ratio ranges from a minimum of $\sim$0.96 to a maximum of $\sim$0.98. Exact values for
each detector are not reported here for simplicity.
    
\begin{figure}[h!]
         \begin{center}
             \includegraphics[width=9cm]{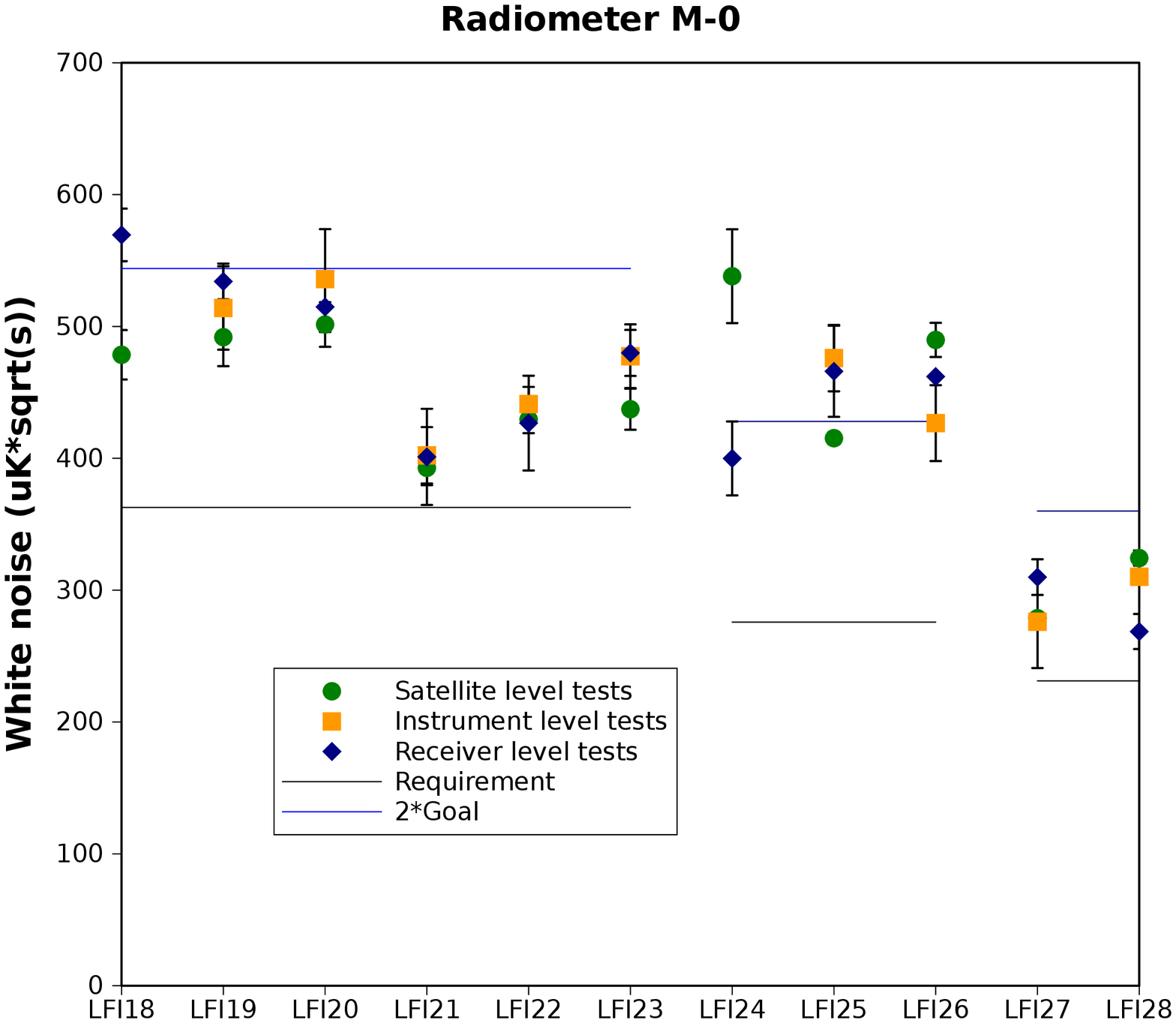} \\
             \mbox{ }\\
             \includegraphics[width=9cm]{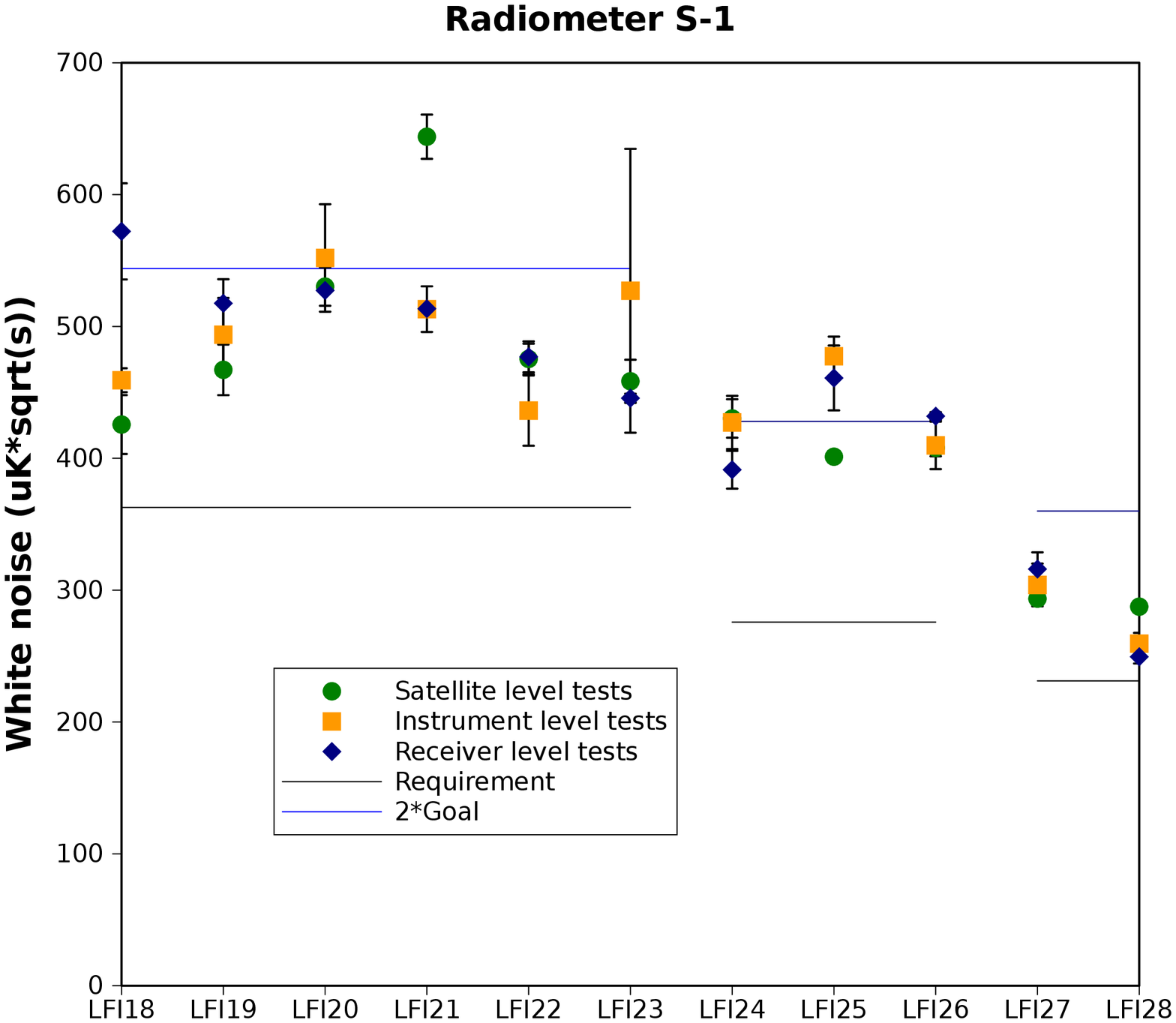} 
         \end{center}
        \caption{Summary of in-flight sensitivities per radiometer estimated from receiver, instrument and satellite
level test campaigns.}
        \label{fig:summary_wn_per_radiometer} 
    \end{figure}

    In Fig.~\ref{fig:summary_wn_per_radiometer} we summarise graphically the in-flight sensitivity estimates from the
three tests. In the following plots the sensitivity values are provided with error bars, with the following meanings:

    \begin{itemize}

        \item error bars in sensitivities estimated from satellite-level data represent the statistical error in the
calibration constants calculated from the various temperature jumps and propagated through the sensitivity formulas.
They represent genuine statistical uncertainties;

        \item error bars in sensitivities estimated from receiver and instrument level tests data represent the
uncertainty coming from the calculation performed according to the two different methods described in
Sect.~\ref{sec:white_noise_eff_bw} and Appendix~\ref{app:white_noise_calibration_extrapolation}. In this case error bars
do not have specific statistical significance, but nevertheless provide an indication of uncertainties in the estimate.

    \end{itemize}
    
    Fig.~\ref{fig:summary_wn_per_radiometer} shows that the in-flight sensitivity lies between the requirement and twice
    the goal levels for the 30 and 70\,GHz receivers, and at about twice the goal for the 44\,GHz receivers. The
    agreement between values extrapolated from the three test campaigns is very good, apart from two noticeable
    outliers, \texttt{LFI21S-1} and \texttt{LFI24M-0}, which showed a higher noise level during satellite level
    tests. Investigation showed that this anomaly was due to incorrect bias voltages on the front-end devices during
    the test. 
    
    After a thorough bias tuning activity conducted during in flight calibration \citep[see][]{2009_LFI_cal_R7}
    a new bias configuration was found that normalised the white noise sensitivity of these two receivers, as expected.
    A full description of the in-flight calibration results and scientific performance will be given in a forthcoming
    dedicated paper.
    
\subsection{Noise stability}
\label{sec:csl_one_over_f}

\begin{figure}[h!]
         \begin{center}
             \includegraphics[width=4.2cm]{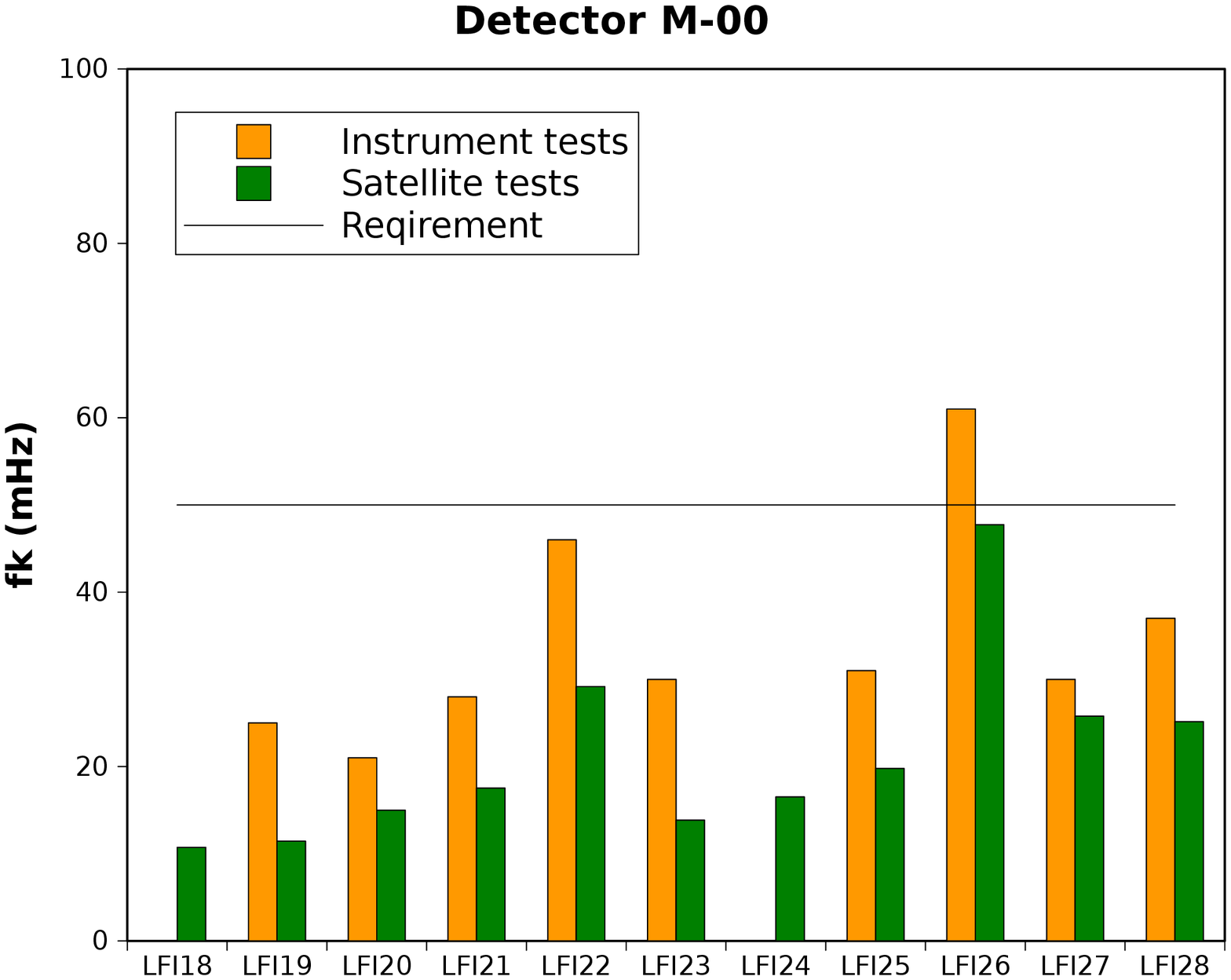} 
             \includegraphics[width=4.2cm]{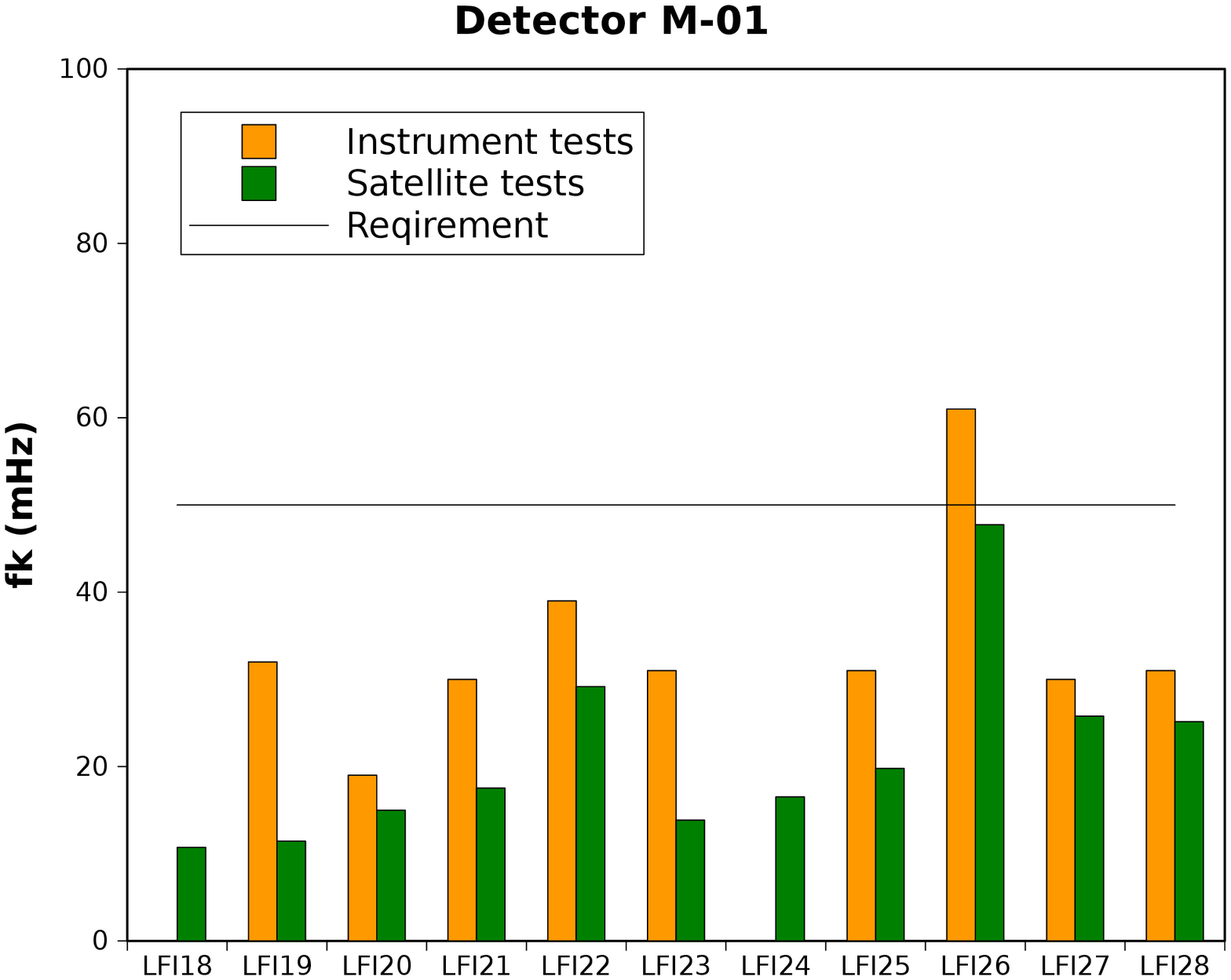} \\
             \vspace{0.5cm}
             \includegraphics[width=4.2cm]{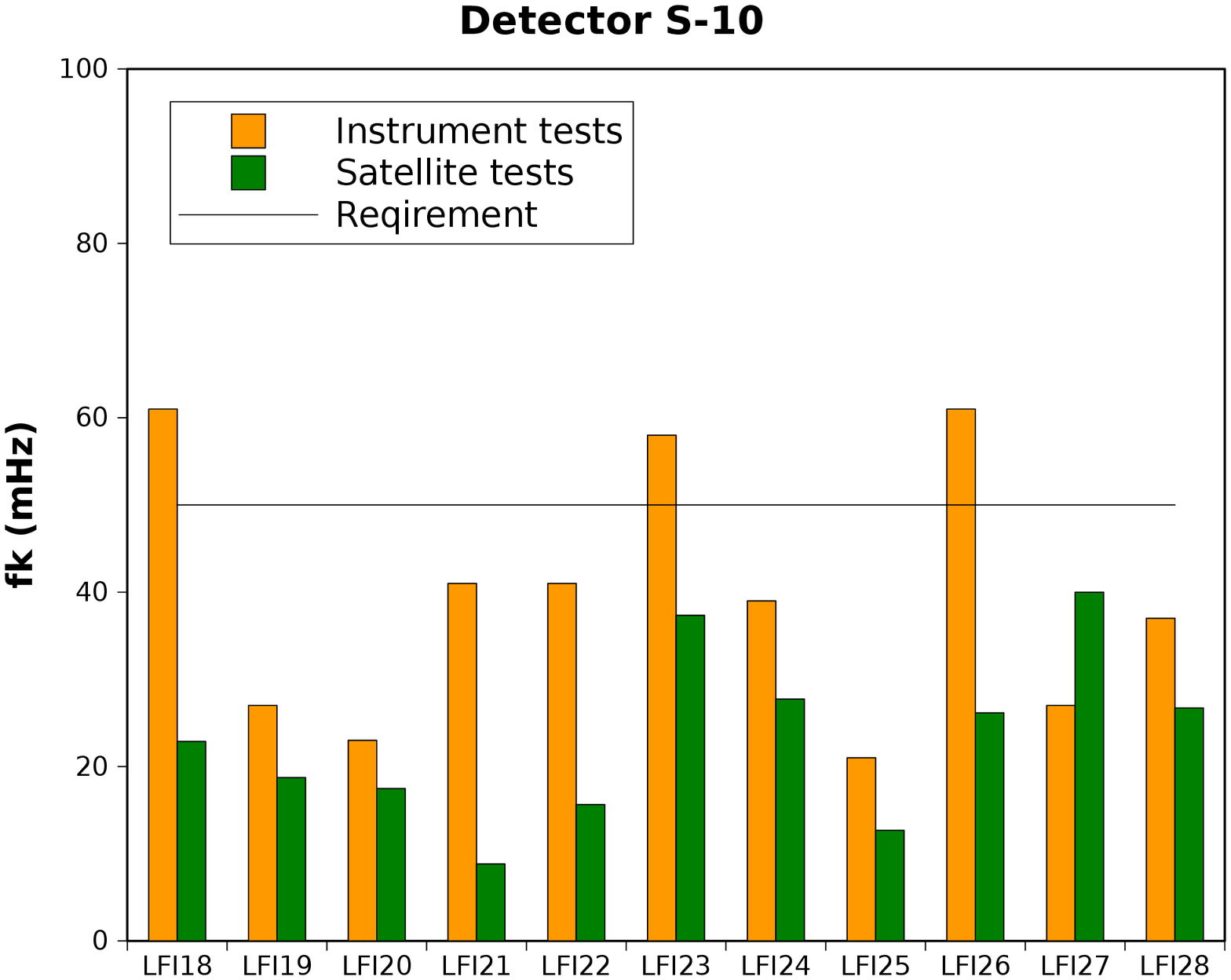} 
             \includegraphics[width=4.2cm]{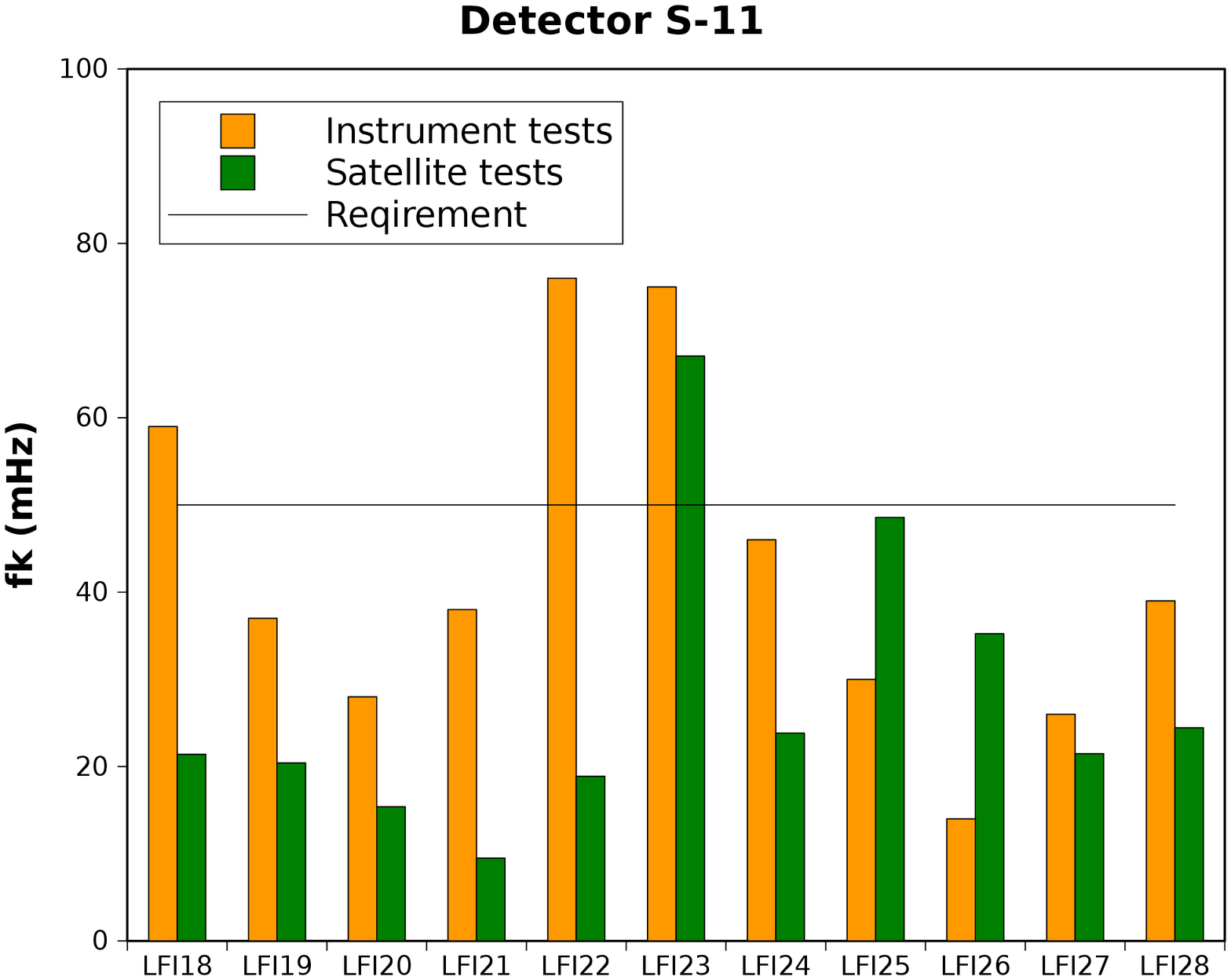} 
         \end{center}
        \caption{Summary of 1/$f$ knee frequencies measured at instrument and satellite levels.}
        \label{fig:summary_fk_per_detector}
    \end{figure}

    Receiver noise stability during satellite-level tests was determined from stable data acquisitions lasting several
hours with the instruments in their tuned and nominal conditions. Fig.~\ref{fig:summary_fk_per_detector} summarises
$1/f$ knee frequencies measured at instrument and satellite levels compared with the 50\,mHz requirement, and shows that
the noise stability of all channels is within requirements, with the single marginal exception of \texttt{LFI23S-11}.
The slope ranged from a minimum of 0.8 to a maximum of 1.7.

    There was a substantial improvement in noise stability during satellite-level tests compared to instrument-level
tests, in some cases with a reduction in knee frequency of more than a factor of 2. This can partly be
explained by the almost perfect signal input balance achieved in the CSL cryo-facility (much less than 1~K compared to
the $\sim$3~K obtained in the instrument cryo-facility). Some of the improvement was also expected because of the much
better thermal stability of the CSL facility. In particular fluctuations of the sky and reference loads in CSL were
about two order of magnitudes less than those in the instrument facility
(see Table~\ref{tab:summary_cryofacility_thermal_performance}).
    Because the highly balanced input achieved in CSL will not be reproduced in flight we expect that the flight knee
frequencies will be slightly higher (although similar) compared to those measured in CSL. 
        
\subsection{Isolation}
\label{sec:csl_isolation}

\begin{figure}[h!]
         \begin{center}
             \includegraphics[width=4.2cm]{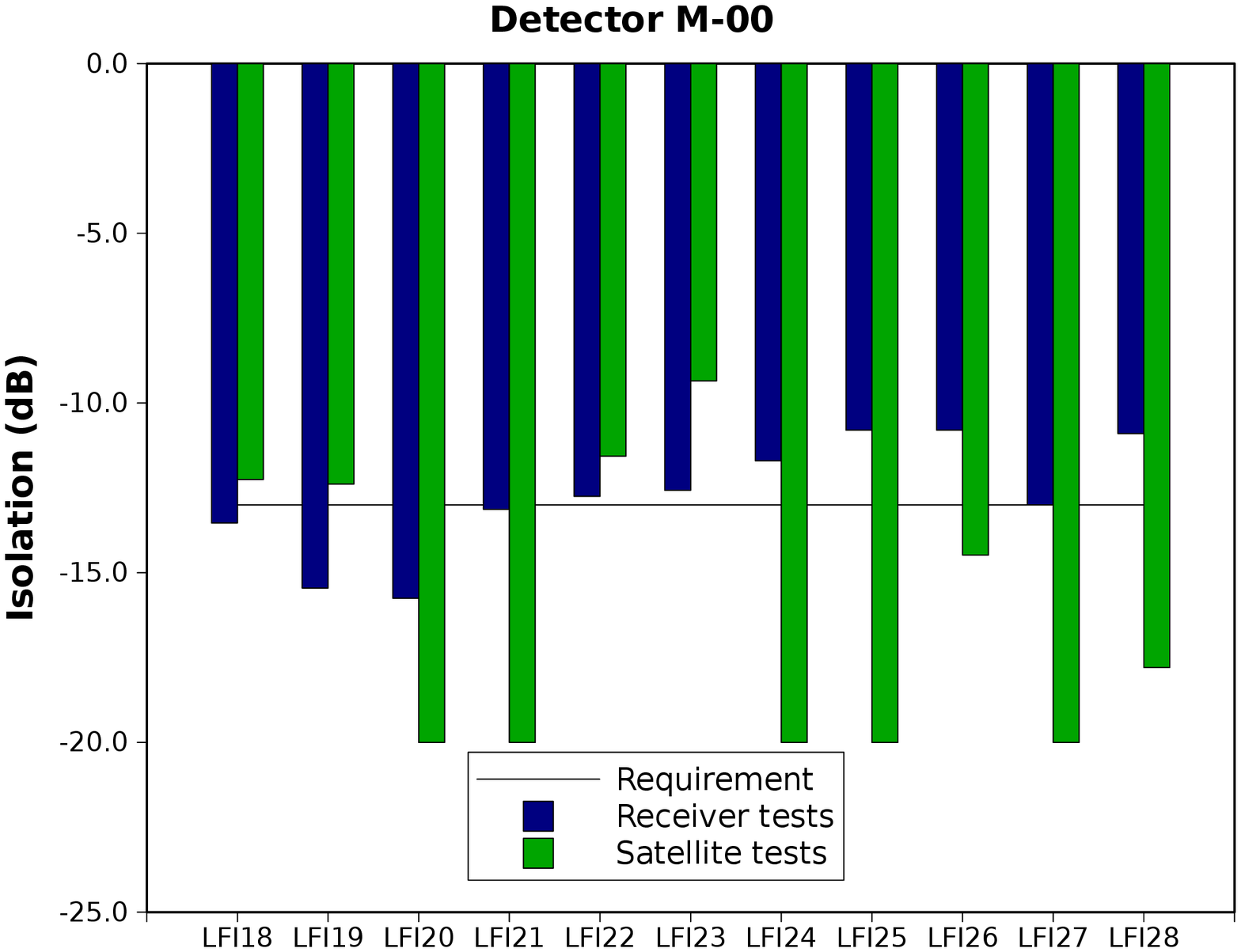} 
             \includegraphics[width=4.2cm]{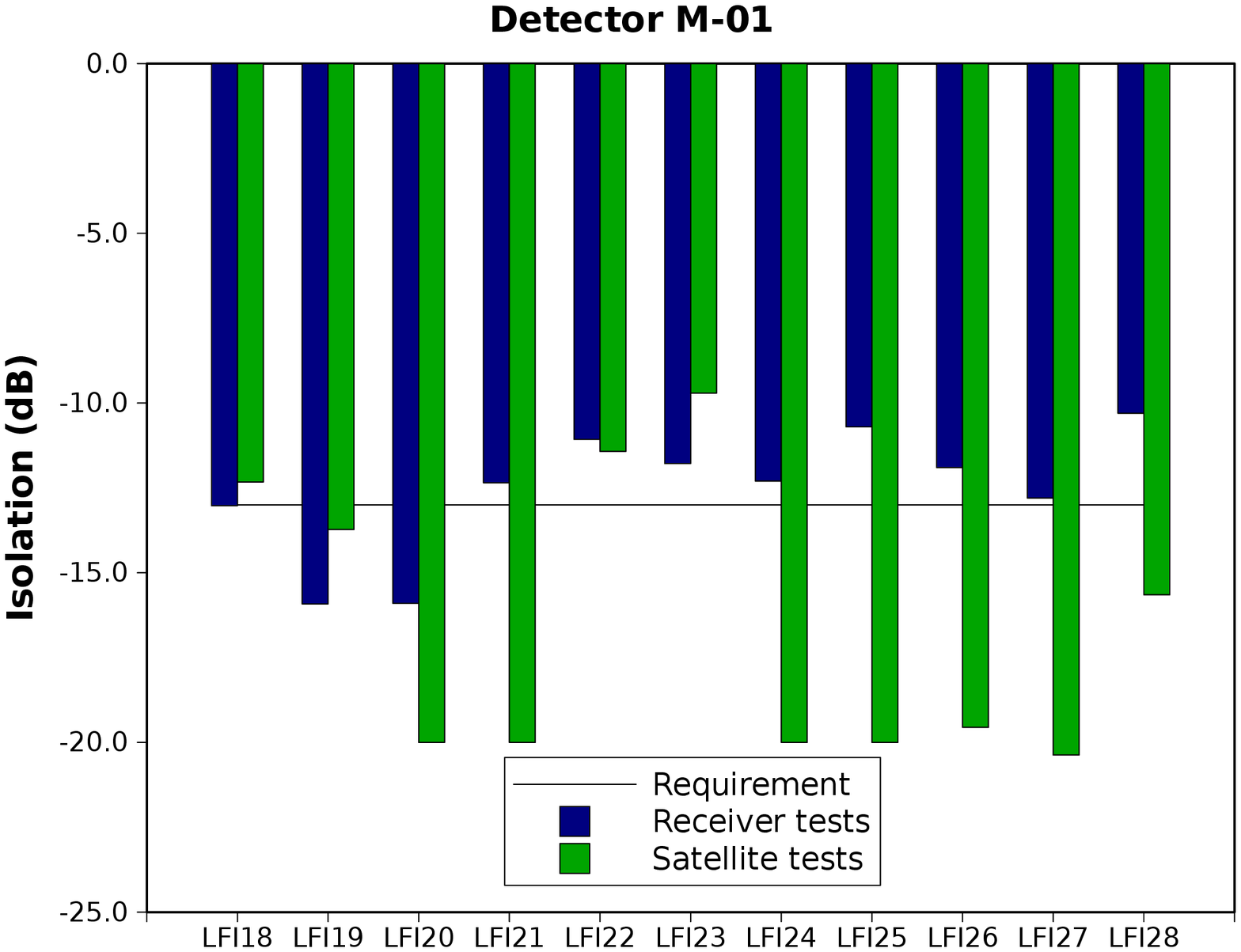} \\
             \vspace{0.5cm}
             \includegraphics[width=4.2cm]{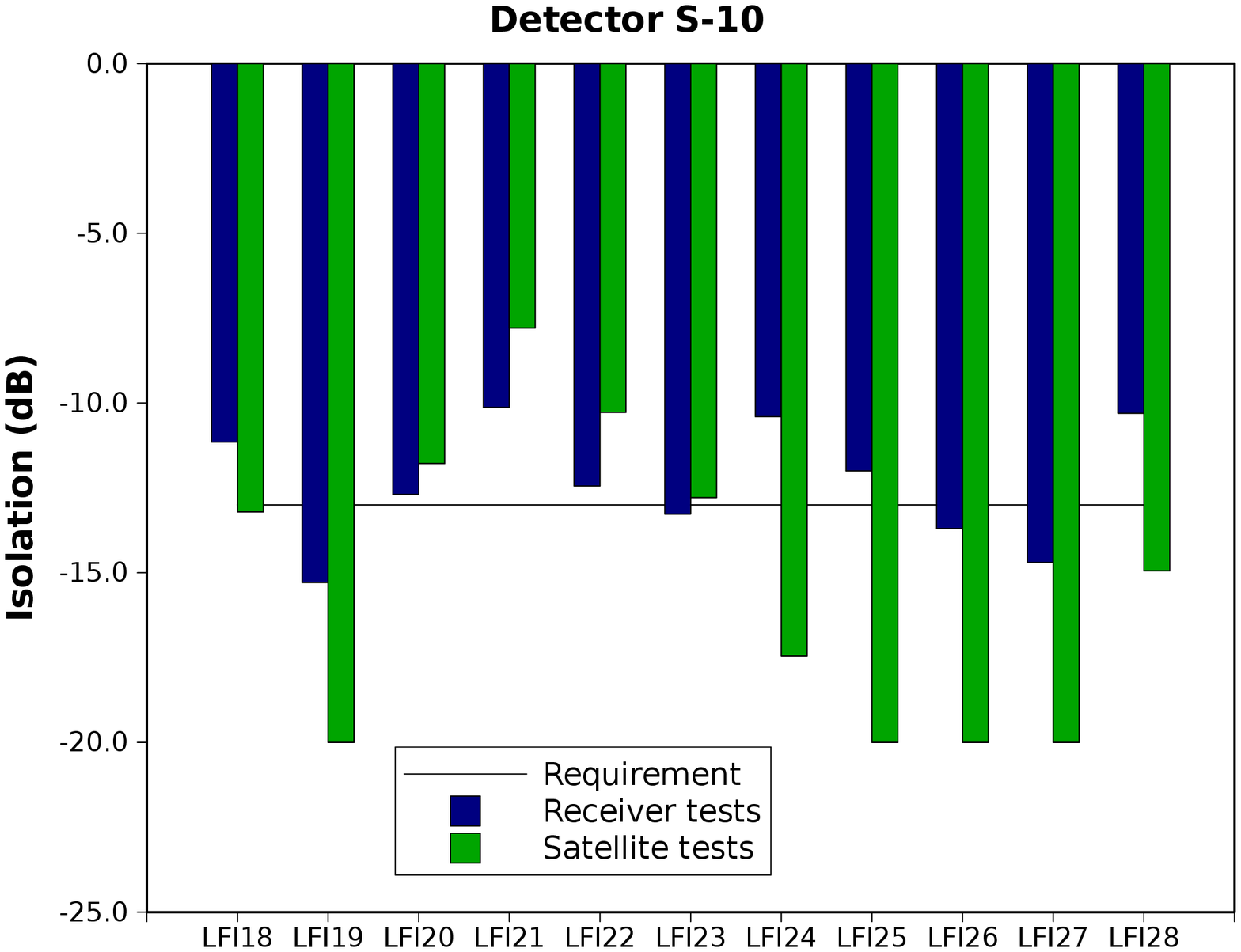} 
             \includegraphics[width=4.2cm]{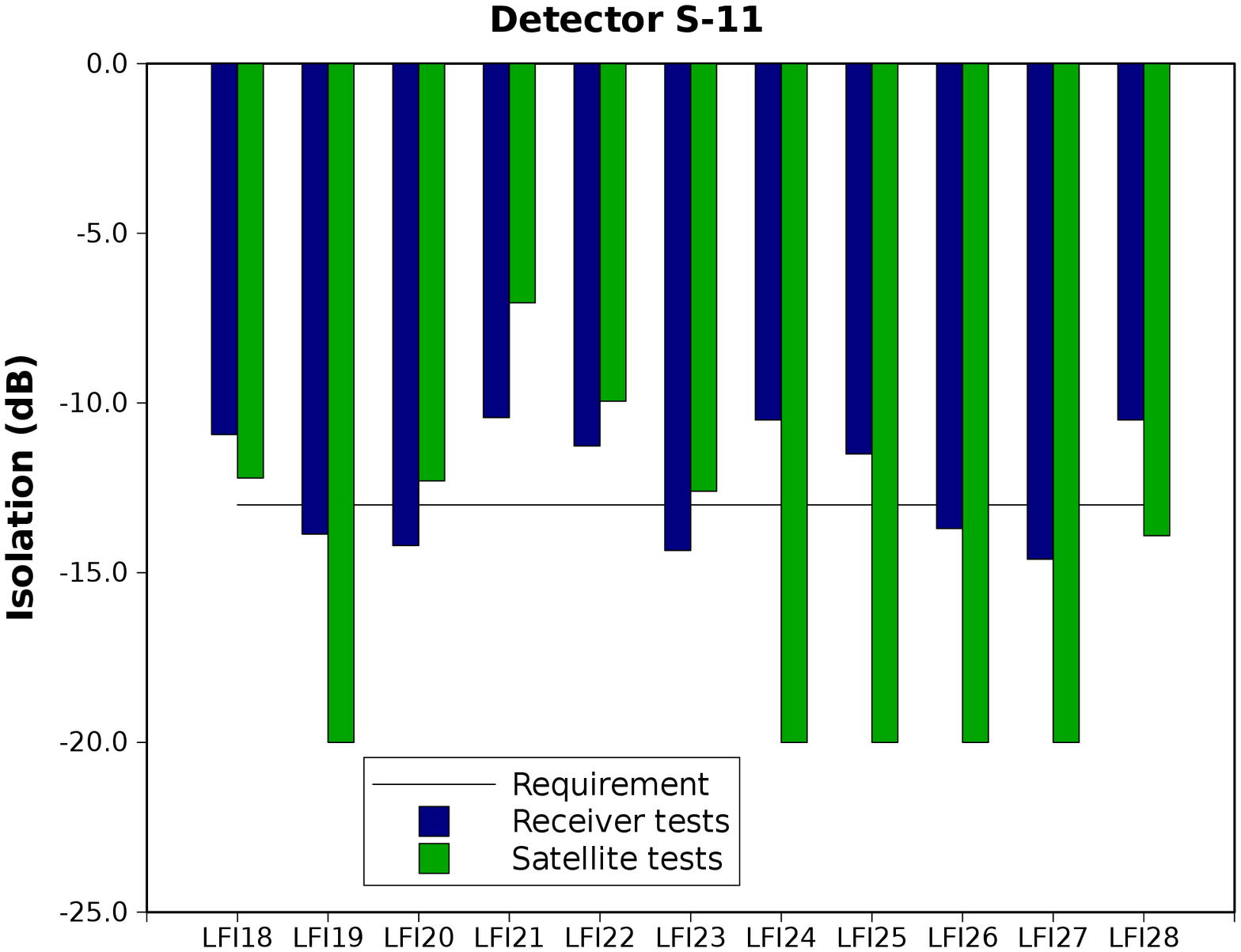} 
         \end{center}
        \caption{Summary of isolation measured at receiver and satellite levels.}
        \label{fig:summary_iso_per_detector}
    \end{figure}

   Isolation (see Eq.~(\ref{eq:isolation_measurement_simple})) was measured during the satellite tests by changing
the reference load temperature by 3.5\,\hbox{K}.  Fig.~\ref{fig:summary_iso_per_detector} compares isolation measured
during receiver- and satellite-level tests.  Several channels exceed the $-13$\,dB requirement; a few are marginally
below. One channel, \texttt{LFI21S-1}, showed poor isolation of only $-7$\,d\hbox{B}.  This result is consis
 with the high value of the calibrated white noise measured for this channel (see Sect.~\ref{sec:csl_white_noise}),
supporting the hypothesis of non-optimal biasing of that channel.

\subsection{Thermal susceptibility}
\label{sec:csl_thermal_suceptibility}
    
As mentioned in Sect.\,4.3, the most important source of temperature fluctuations in the LFI focal plane is the sorption
cooler.  The satellite-level test provided the first opportunity to measure the performance of the full Planck thermal
system.  Fuctuations at the interface between the sorption cooler and the LFI were measured to be about 100\,mK
peak-to-peak.  Using methods described in \citep{mennella02}, we calculate that the effect of these fluctuations will be
less than 1\,$\mu$K per pixel in the maps, in line with the scientific requirements outlined in \citet{2009_LFI_cal_M2}.


\section{Conclusions}
\label{sec:conclusions}

The LFI was integrated and tested in thermo-vacuum conditions at the Thales Alenia Space Italia laboratories, located
in Vimodrone (Milano), during the summer of 2006. The test goals were a wide characterisation and calibration of the
instrument, ranging from functionality to scientific performance assessment.

The LFI was fully functional, apart from two failed components in \texttt{LFI18M-0} and \texttt{LFI24M-0} that have
been fixed (one replaced and the other repaired) after the cryogenic test campaign, recovering full functionality.
 
Measured instrument parameters are consistent with measurements performed on indivudual receivers.  In particular, the
LFI shows excellent $1/f$ stability and rejection of instrumental systematic effects. Although the very ambitious
sensitivity goals have not been fully met, the measured performance makes LFI the most sensitive instrument of its kind,
a factor of 2 to 3 better than WMAP\footnote{Calculated on the final resolution element per unit integration time} at
the same frequencies. In particular at 70\,GHz, near the minimum of the foreground emission for both temperature and
polarisation anisotropy, the combination of sensitivity and angular resolution of LFI will provide a clean
reconstruction of the temperature power spectrum up to $\ell\sim 1400$ \citep{2009_LFI_cal_M1}.

After the instrument test campaign, the LFI was integrated with the HFI and the satellite. Between June and August
2008, Planck was tested at the Centre Spatiale de Li\`ege in flight-representative, thermo-vacuum conditions, and showed
to be fully functional.

Planck was launched on May 14$^{\rm th}$ from the Guyane Space Centre in Kourou and has reached its observation point,
L2.  In-flight testing and calibration is underway,  and will provide the final instrument tuning and scientific
performance assessment. After 17 years, Planck is nearly ready to start recording the first light in the Universe.


\begin{acknowledgements}
     The Planck-LFI project is developed by an Interntional Consortium lead by Italy and involving Canada, Finland, Germany, Norway, Spain, Switzerland, UK, USA. The Italian contribution to Planck is supported  by the Italian Space Agency (ASI). The work in this paper has been supported by in the framework of the ASI-E2 phase of the Planck contract. The US Planck Project is supported by the NASA Science Mission Directorate. In Finland, the Planck LFI 70 GHz work was supported by the Finnish Funding Agency for Technology and Innovation (Tekes).
\end{acknowledgements}


%
\clearpage 

    \begin{table*}
     
      \caption{Best fit parameters obtained from the non linear fit of data acquired during instrument-level tests. 
       Notice that the linearity factor was obtained by constraining it to $\pm 1$\% around the
       value found during calibration of individual receivers.\citep[see][]{2009_LFI_cal_R4}}
      \label{tab:tnoise_gain_lin_results}
      \begin{center}  
        \begin{tabular}{cccccc}
        \mbox{}\\
        
        \hline   
        \hline
        Rec. ID & Param. & M-00 & M-01 & S-10 & S-11 \\
        \hline   
        \mbox{}\\
                        & $b$          & $\ldots$  & $\ldots$  & $\lesssim 10^{-3}$  & $\lesssim 10^{-3}$  \\
        \textbf{LFI18} & $G_0$ (V/K)  & $\ldots$ & $\ldots$ & 0.026 & 0.022 \\
        \vspace{.2cm}
                        & $T_{\rm noise}$ (K)  & $\ldots$ & $\ldots$ & 37.4 & 40.5 \\
        
                        & $b$  & $\lesssim 10^{-3}$   & $\lesssim 10^{-3}$   & $\lesssim 10^{-3}$   & $\lesssim 10^{-3}$
  \\
        \textbf{LFI19} & $G_0$ (V/K)  & 0.020 & 0.021 & 0.016 & 0.018 \\
        \vspace{.2cm}
                        & $T_{\rm noise}$ (K)  & 39.8 & 38.7 & 37.5 & 40.0 \\
        
                        & $b$  & $\lesssim 10^{-3}$  & $\lesssim 10^{-3}$   & $\lesssim 10^{-3}$   & $\lesssim 10^{-3}$ 
 \\
        \textbf{LFI20} & $G_0$ (V/K)  & 0.019 & 0.018 & 0.025 & 0.025 \\
        \vspace{.2cm}
                        & $T_{\rm noise}$ (K)  & 42.3 & 42.2 & 43.9 & 43.0 \\
        
                        & $b$  & $\lesssim 10^{-3}$  & $\lesssim 10^{-3}$  & $\lesssim 10^{-3}$  & $\lesssim 10^{-3}$  
\\
        \textbf{LFI21} & $G_0$ (V/K)  & 0.025 & 0.023 & 0.016 & 0.014 \\
        \vspace{.2cm}
                        & $T_{\rm noise}$ (K)  & 31.9 & 34.6 & 43.3 & 45.9 \\
        
                        & $b$  & $\lesssim 10^{-3}$  & $\lesssim 10^{-3}$   & $\lesssim 10^{-3}$  & $\lesssim 10^{-3}$ 
\\
        \textbf{LFI22} & $G_0$ (V/K)  & 0.011 & 0.012 & 0.014 & 0.016 \\
        \vspace{.2cm}
                        & $T_{\rm noise}$ (K)  & 40.5 & 38.9 & 40.8 & 43.5 \\
        
                        & $b$  & $\lesssim 10^{-3}$   & $\lesssim 10^{-3}$   & $\lesssim 10^{-3}$   & $\lesssim 10^{-3}$
  \\
        \textbf{LFI23} & $G_0$ (V/K)  & 0.025 & 0.029 & 0.014 & 0.007 \\
        \vspace{.2cm}
                        & $T_{\rm noise}$ (K)  & 40.6 & 39.2 & 50.3 & 54.2 \\
        
                        & $b$  & $\ldots$ & $\ldots$ & 1.43 & 1.43  \\
        \textbf{LFI24} & $G_0$ (V/K)  & $\ldots$ & $\ldots$ &  0.005 &   0.005  \\
        \vspace{.2cm}
                        & $T_{\rm noise}$ (K)  & $\ldots$ & $\ldots$ &  19.7 & 19.9 \\
    
                        & $b$  &  1.21 &  1.16 &  0.79 & 1.00 \\
        \textbf{LFI25} & $G_0$ (V/K)  &  0.008 & 0.008 &   0.007 &  0.007 \\
        \vspace{.2cm}
                        & $T_{\rm noise}$ (K)  & 19.7 &  19.7 &  20.5 & 20.2 \\
        
                        & $b$  &  1.07 & 1.40 & 0.93 & 1.21 \\
        \textbf{LFI26} & $G_0$ (V/K)  &  0.005 &  0.006  &    0.007 & 0.007\\
        \vspace{.2cm}
                        & $T_{\rm noise}$ (K)  & 20.2 &  19.1 &  18.5 &  18.1 \\
        
                        & $b$  &0.12 & 0.12 & 0.12 & 0.14 \\
        \textbf{LFI27} & $G_0$ (V/K)  & 0.074 &    0.081 &   0.070 &   0.058 \\
        \vspace{.2cm}
                        & $T_{\rm noise}$ (K)  & 13.3 &  13.1 &  14.3  &  13.7 \\
        
                        & $b$  & 0.19 &  0.16 & 0.19 & 0.19 \\
        \textbf{LFI28} & $G_0$ (V/K)  & 0.076 &   0.103 &   0.071 &    0.061 \\
        \vspace{.2cm}
                        & $T_{\rm noise}$ (K)  & 11.7 &  11.3 &  10.9 &  10.8 \\      
        \hline      

      \end{tabular}
      \end{center}      
    \end{table*}

      \begin{table*}
         \caption{
            Measured white noise spectral densities (in $\mu$V$/\sqrt{\rm Hz}$)
            before and after quantisation and compression for 
            all detectors. No values are given for \texttt{LFI26S-11}, for which quantisation
            and compression parameters were set to wrong values because of a problem in the software optimisation
            procedure that was identified and solved after the calibration campaign.
         }
         \label{tab:white_noise_before_after_compression}

         \begin{center}
            \begin{tabular}{| c | c c c | c c c | c c c | c c c |}
               \hline
               \hline
                              & &\textbf{M-00}  &       &       &\textbf{M-01}  &       &       &\textbf{S-10}  &      
&       &\textbf{S-11}  &\\
         \cline{2-13}
         &$\sigma$      &$\sigma_q$     &$\Delta$       
         &$\sigma$      &$\sigma_q$     &$\Delta$       
         &$\sigma$      &$\sigma_q$     &$\Delta$       
         &$\sigma$      &$\sigma_q$     &$\Delta$       \\
           \hline
               \textbf{70 GHz} & & &  &  &  &  & & &  &  &  & \\
           LFI18        &$\ldots$       &$\ldots$       &$\ldots$       &$\ldots$       &$\ldots$       &$\ldots$      
&38.93  &39.22  &0.74\% &31.07  &31.39  &1.02\%\\
               LFI19    &33.50  &33.68  &0.55\% &34.00  &34.13  &0.39\% &25.68  &25.85  &0.63\% &27.48  &27.67 
&0.71\%\\
               LFI20    &31.08  &31.17  &0.31\% &31.20  &31.37  &0.54\% &44.77  &45.14  &0.83\% &41.95  &42.23 
&0.67\%\\
               LFI21    &33.77  &33.94  &0.51\% &32.27  &32.39  &0.35\% &26.50  &26.67  &0.62\% &25.63  &25.86 
&0.87\%\\
               LFI22    &17.03  &17.15  &0.67\% &19.29  &19.41  &0.61\% &20.99  &21.05  &0.28\% &23.94  &24.06 
&0.49\%\\
               LFI23    &37.84  &38.01  &0.44\% &41.00  &41.25  &0.61\% &23.76  &24.01  &1.04\% &12.15  &12.19 
&0.36\%\\
           \hline   
           \textbf{44 GHz} & & &  &  &  &  & & &  &  &  & \\
               LFI24    &$\ldots$       &$\ldots$       &$\ldots$       &$\ldots$       &$\ldots$       &$\ldots$      
&5.95   &5.97   &0.25\% &5.32   &5.35   &0.45\%\\
               LFI25    &7.50   &7.54   &0.50\% &7.53   &7.55   &0.30\% &9.34   &9.37   &0.35\% &6.93   &6.96  
&0.43\%\\
               LFI26    &6.04   &6.06   &0.32\% &6.18   &6.20   &0.31\% &8.81   &8.84   &0.28\% &$\ldots$      
&$\ldots$       &$\ldots$       \\
           \hline   
           \textbf{30 GHz} & & &  &  &  &  & & &  &  &  & \\
           LFI27        &62.34  &62.67  &0.52\% &65.62  &65.97  &0.53\% &56.19  &56.40  &0.37\% &52.48  &52.59 
&0.22\%\\
               LFI28    &52.96  &53.27  &0.59\% &68.34  &68.58  &0.34\% &46.77  &46.94  &0.35\% &44.15  &44.24 
&0.20\%\\
               \hline
            \end{tabular}
         \end{center}
      \end{table*}

\clearpage 

\appendix

\section{LFI receiver and channel naming convention}
\label{app:naming_convention}

\begin{figure}[h!]
    \begin{center}
       \includegraphics[width=7.5cm]{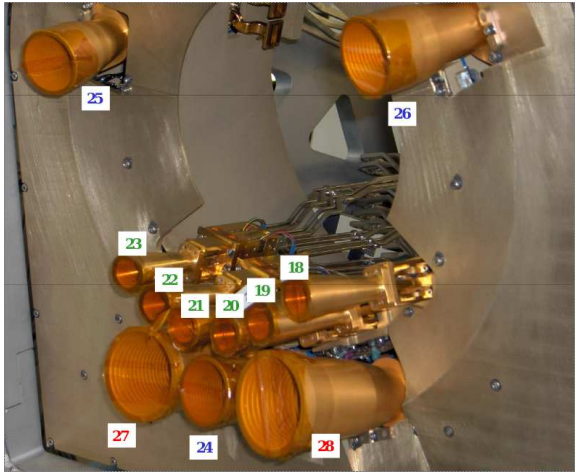}
    \end{center}
   \caption{
        Feed horns in the LFI focal plane. Each feed horn is tagged by a label running fro LFI18 to LFI28. LFI18 through
LFI23 are 70~GHz receivers, LFI24 through LFI26 are 44~GHz receivers and LFI27, LFI28 are 30~GHz receivers.
}
   \label{fig:LFI_focal_plane}
\end{figure}

The various receivers are labelled LFI18 to LFI28, as shown in Fig.~\ref{fig:LFI_focal_plane}. The radiometers connectd
to the two OMT arms are labelled  M-0 (``main'' OMT arm) and S-1 (``side'' OMT arm), while the two output detectors from
each radiometer are be labelled as 0 and 1. Therefore \texttt{LFI18S-10}, for example, refers to detector 0 of the side
arm of  receiver LFI18, and \texttt{LFI24M-01} refers to detector 1 of the main arm of receiver LFI24.

\section{Receiver isolation: definition, scientific requirements and measurements}
\label{app:isolation}
      
\subsection{Definition and requirement.}
In Sect.~\ref{sec:lfi_overview} it is shown that the output of the LFI pseudo-correlation
receivers is a sequence of sky and reference load signals alternating at twice the phase switch frequency.       If the
pseudo-correlator is not ideal, the separation after the second hybrid is not perfect and a certain level of mixing
between the two signals will be present in the output. Typical limitations on isolation are (i) imperfect hybrid phase
matching, (ii) front-end gain amplitude mismatch, and (iii) mismatch in the insertion loss in the two switch states
\citep{seiffert02}.

      A more general relationship representing the receiver power output can be written as:

      \begin{eqnarray}
        p_{\rm out} &=& a G_{\rm tot} k \beta \left[
        (1-\epsilon)T_{\rm sky} + \epsilon T_{\rm ref} + 
            T_{\rm noise} + \right.\nonumber\\
            &-& \left.  r \left(
        (1-\epsilon)T_{\rm ref} + \epsilon T_{\rm sky} + T_{\rm noise} \right)
        \right]
        \label{eq:isolation_power_output}
      \end{eqnarray}
      where the parameters $\epsilon$ represents the degree of mixing or, in other words, deviation from ideal
isolation. 

      Let us now imagine the receiver scanning the sky and therefore measuring a variation in the sky signal given by
the   CMB, $\Delta T_{\rm CMB}$. If we define $r= \frac{T_{\rm sky} + T_{\rm noise}}{T_{\rm ref} + T_{\rm noise}}$ and
develop Eq.~(\ref{eq:isolation_power_output}) in series up to the first order in $\epsilon$ we see that the differential
power output is proportional to:

      \begin{equation}
         p_{\rm out} \propto \Delta T_{\rm CMB}\left(1 - \delta_{\rm iso}\right)
      \end{equation}
      where $\delta_{\rm iso} = \frac{2 T_{\rm noise} + T_{\rm sky} + T_{\rm ref}}{T_{\rm noise} + T_{\rm
ref}}\epsilon$, which provides a useful relationship to estimate the requirement on the isolation, $\epsilon_{\rm max}$
given an acceptable level of $\delta_{\rm iso}^{\rm max}$.

      If we assume 10\% (corresponding to $\delta_{\rm iso}^{\rm max}\sim 0.1$) as the maximum acceptable loss in the
CMB signal due to imperfect isolation and consider typical values for the LFI receivers ($T_{\rm ref}= 4.5$~K and
$T_{\rm noise}$ ranging from 10 to 30\,K), we find $\epsilon_{\rm max}= 0.05$ equivalent to $-13$~dB, which corresponds
to the requirement for LFI receivers.

      \subsection{Measurement.}

      If $\Delta V_{\rm sky}$ and $\Delta V_{\rm ref}$ are the voltage output variations induced by $\Delta T = T_2 -
T_1$, then it is easy to see from Eq.~(\ref{eq:isolation_power_output}) (with the approximation $(1-\epsilon)\simeq 1$)
that:

      \begin{equation}
         \epsilon \simeq \frac{\Delta V_{\rm ref}}{\Delta V_{\rm sky} + \Delta V_{\rm ref}}.
         \label{eq:isolation_measurement_simple}
      \end{equation}

      If the reference load temperature is not perfectly stable but varies by an amount $\Delta T_{\rm ref}$ during the
measurement, this can be corrected at first order if we know the photometric constant $G_0$. In this case
Eq.~(\ref{eq:isolation_measurement_simple}) becomes:

      \begin{equation}
         \epsilon \simeq \frac{\Delta V_{\rm ref} - G_0 \Delta T_{\rm ref}}{\Delta V_{\rm sky} + \Delta V_{\rm
ref}-G_0\Delta T_{\rm ref}}.
         \label{eq:isolation_measurement_complete}         
      \end{equation}

      Measuring the isolation accurately, however, is generally difficult and requires a very stable environment. In
fact, any change in $\Delta V_{\rm ref}$ caused by other systematic fluctuations (e.g., temperature fluctuations, $1/f$
noise fluctuations) will impact the isolation measurement causing an over- or under-estimation depending on the sign of
the effect.

      To estimate the accuracy in our isolation measurements, we have first calculated the uncertainty caused by a
systematic error in the reference load voltage output, $\Delta V_{\rm ref}^{\rm sys}$. If we substitute in
Eq.~(\ref{eq:isolation_measurement_complete}) $\Delta V_{\rm ref}$ with $\Delta V_{\rm ref} \pm \Delta V_{\rm ref}^{\rm
sys}$ and develop at first order in $\Delta V_{\rm ref}^{\rm sys}$, we obtain

      \begin{equation}
         \epsilon \sim \epsilon_0 \mp \frac{\Delta V_{\rm sky}}
{\Delta V_{\rm sky} + \Delta V_{\rm ref} - G_0 \Delta T_{\rm ref}}\Delta V_{\rm ref}^{\rm sys}
            \equiv \epsilon_0 \mp \delta \epsilon,
         \label{eq:isolation_with_error}
      \end{equation}
      where we indicate with $\epsilon_0$ the isolation given by Eq.~(\ref{eq:isolation_measurement_complete}).

      We estimated $\delta \epsilon$ in our measurement conditions. Because the three temperature steps were implemented
in about one day we have evaluated the total power signal stability on this timescale from a long-duration acquisition
in which the instrument was left running undisturbed for about two days. For each detector datastream we have first
removed spurious thermal fluctuations by correlation analysis with temperature sensor data   then we calculated the
peak-to-peak variation in the reference load datastream.

\section{Calculation of noise effective bandwidth}
\label{app:noise_eff_bw_calculation}

   The well-known radiometer equation applied to the output of a single diode in the Planck LFI receivers links the
white noise sensitivity to sky and noise temperatures and the receiver bandwidth. It reads \citep{seiffert02}:

   \begin{equation}
      \delta T_{\rm rms} = 2\frac{T_{\rm sky}+T_{\rm noise}}{\sqrt{\beta}}.
      \label{eq:radiometer_equation}
   \end{equation}

   In the case of linear response, i.e. if $V_{\rm out} = G\times(T_{\rm sky}+T_{\rm noise})$ (where $G$ represents the
photometric calibration constant) we can write Eq.~(\ref{eq:radiometer_equation}) in its most useful uncalibrated form:

   \begin{equation}
      \delta V_{\rm rms} = 2\frac{V_{\rm out}}{\sqrt{\beta}},
      \label{eq:radiometer_equation_uncalibrated}      
   \end{equation}
   which is commonly used to estimate the receiver bandwidth, $\beta$, from a simple measurement of the receiver DC
output and white noise level, i.e.:

   \begin{equation}
      \tilde\beta = 4\left(\frac{V_{\rm out}}{\delta V_{\rm rms}}\right)^2.
      \label{eq:noise_effective_bandwidth}
   \end{equation}

   If the response is linear and if the noise is purely radiometric (i.e. all the additive noise from back end
electronics is negligible and if there are no non-thermal noise inputs from the source) then $\tilde \beta$ is
equivalent to the receiver bandwidth, i.e. 

   \begin{equation}
      \tilde \beta \equiv \beta = 4\left(\frac{T_{\rm sky}+T_{\rm noise}}{\delta T_{\rm rms}}\right)^2.
      \label{eq:bandwidths_equivalence}
   \end{equation}

   Conversely, if the receiver output is compressed, from Eq.~(\ref{eq:nonlinear_response}) we have that:
   
   \begin{equation}
      \delta V_{\rm rms} = \frac{\partial V_{\rm out}}{\partial T_{\rm in}}\delta T_{\rm rms}.
      \label{eq:delta_vrms}
   \end{equation}

   By combining Eqs.~ (\ref{eq:nonlinear_response}), (\ref{eq:noise_effective_bandwidth}) and  (\ref{eq:delta_vrms}) we
find:
   
   \begin{eqnarray}
      \tilde \beta &=& 4\left(\frac{T_{\rm sky}+T_{\rm noise}}{\delta T_{\rm rms}}\right)^2 
         \left[ 1 + b\, G_0(T_{\rm sky}+T_{\rm noise})\right]^2 \equiv \nonumber\\
      &\equiv& \beta  \left[ 1 + b\, G_0(T_{\rm sky}+T_{\rm noise})\right]^2,
      \label{eq:bandwidth_compressed}
   \end{eqnarray}
   which shows that $\tilde \beta$ is an overestimate of the ``optical'' bandwidth unless the non linearity parameter
$b$ is very small.

\section{White noise sensitivity calibration and extrapolation to flight conditions}
\label{app:white_noise_calibration_extrapolation}

   In this appendix we detail the calculation needed to convert the uncalibrated white noise sensitivity measured on the
ground to the expected calibrated sensitivity in flight conditions. The calculation starts from the general radiometric
output model in Eq.~(\ref{eq:nonlinear_response}), which can be written in the following form:

   \begin{equation}
      T_{\rm out}(V_{\rm in}) = \-T_{\rm noise}-\frac{V_{\rm in}}{G_0 (b\, V_{\rm in}-1)}
   \end{equation}
   
   Our starting point is the the raw datum, that is a couple of uncalibrated white noise levels for the two detectors in
a radiometer measured with the sky load at a temperature $T_{\rm {sky-load}}$ and the front end unit at physical
temperature $T_{\rm test}$. 

   From the measured uncalibrated white noise level in Volt\,s$^{1/2}$, we want to derive a calibrated white noise level
extrapolated to input temperature equal to $T_{\rm sky}$ and with the front end unit at a temperature of $T_{\rm nom}$.
This is done in three steps:

   \begin{enumerate}
      \item extrapolation to nominal front-end unit temperature;
      \item extrapolation to nominal input sky temperature;
      \item calibration in units of  K\,s$^{1/2}$.
   \end{enumerate}

   In the following sections we will describe in detail the calculations underlying each step.

\subsection{Step 1---extrapolate uncalibrated noise to 
   nominal front end unit temperature}

   This is a non-trivial step to be performed if we want to consider all the elements in the extrapolation. Here we
focus on a zero-order approximation based on the following assumptions:

   \begin{enumerate}
      \item the radiometer noise temperature is dominated by the front-end noise temperature, such that $T_{\rm noise}
\sim T_{\rm noise}^{\rm FE}$;
      \item we neglect any effect on the noise temperature given by resistive losses of the front-end passive
components;
      \item we assume the variation of $T_{\rm noise}^{\rm FE}$ to be linear in $T_{\rm phys}$.
   \end{enumerate}

   Under these assumptions the receiver noise temperature at nominal front-end temperature can be written as

   \begin{equation}
      T_{\rm noise}(T_{\rm nom}) = T_{\rm noise}(T_{\rm test}) + 
      \frac{\partial T_{\rm noise}^{\rm FE}}{\partial T_{\rm phys}}\Delta
      T_{\rm phys}, 
      \label{eq:noise_temperature_extrapolation_to_nominal_fem_temperature}
   \end{equation}
   where $\Delta T_{\rm phys} = T_{\rm nom} - T_{\rm test}$. A similar but slightly different relationship yields for
the gain factor $G_0$. In fact let us consider that $G_0 = {\rm const}\times G^{\rm FE}\, G^{\rm BE}$, and that we can
write $G^{\rm FE}(T_{\rm nom}) = G^{\rm FE}(T_{\rm test})(1+\delta)$, where $      \delta = \frac{1}{G^{\rm FE} (T_{\rm
test})} \frac{\partial G^{\rm FE}}{\partial T_{\rm phys}} \Delta T_{\rm phys} = \frac{\ln(10)}{10}\frac{\partial G^{\rm
FE}({\rm dB})}{\partial T_{\rm phys}}\Delta T_{\rm phys}$, i.e.,

   \begin{equation}
      G_0(T_{\rm nom}) = G_0(T_{\rm test}) (1+\delta).
      \label{eq:gain_extrapolation_to_nominal_fem_temperature}
   \end{equation}

   From the radiometer equation we have that 
   $\sigma \propto (T_{\rm in} + T_{\rm noise})$, from which we can write

   \begin{eqnarray}
      \sigma(T_{\rm nom}) \equiv \sigma^{\rm nom} &=& \sigma(T_{\rm test})
      \frac{(T_{\rm in} + T_{\rm noise}(T_{\rm nominal}))}
         {(T_{\rm in} + T_{\rm noise}(T_{\rm test}))} =\nonumber \\ 
         &=& \sigma(T_{\rm test})(1+\eta),
      \label{eq:white_noise_extrapolation_to_nominal_fem_temperature}
   \end{eqnarray}
   where
   \begin{equation}
      \eta = \frac{\partial T_{\rm noise}^{\rm FE}}
      {\partial T_{\rm phys}}
      \left[ (T_{\rm in} + T_{\rm noise}(T_{\rm test}))\right]^{-1}
      \Delta T_{\rm phys}.
   \end{equation}

\subsection{Step 2---extrapolate uncalibrated noise to $T_{sky}$}

From this point we will consider quantities ($T_{\rm noise}$, white noise level, and $G_0$) already extrapolated to the
nominal front end temperature using Eqs.
(\ref{eq:noise_temperature_extrapolation_to_nominal_fem_temperature}), 
(\ref{eq:gain_extrapolation_to_nominal_fem_temperature}) and 
(\ref{eq:white_noise_extrapolation_to_nominal_fem_temperature}).
Therefore we will now omit the superscript ``nom'' so that, for example, $\sigma \equiv \sigma^{\rm nom}$.

Let us start from the radiometer equation in which, for each detector, the white noise spectral density is given by

\begin{equation}
   \delta T_{\rm rms} = 2\frac{T_{\rm in}+T_{\rm noise}}{\sqrt{\beta}}.
   \label{eq:single_diode_radiometer_equation}
\end{equation}

Now we want to find a similar relationship for the uncalibrated white noise spectral density linking $\delta V_{\rm
rms}$ to $V_{\rm out}$. 
We start from Eq.~(\ref{eq:delta_vrms}) and calculate the derivative of $V_{\rm out}$ using
Eq.~(\ref{eq:nonlinear_response}) and $\delta T_{\rm rms}$ from Eq.~(\ref{eq:single_diode_radiometer_equation}). We
obtain

\begin{equation}
   \sigma=\frac{V_{\text{out}}}{\sqrt{\beta }}\left[1+ 
   b\, G_0\left(T_{\text{in}}+T_n\right)\right]^{-1},
   \label{eq:white_noise}
\end{equation}
%
%
where $\beta$ is the bandwidth and $V_{\rm out}$ is the DC voltage output of the receiver. Considering the two input
temperatures $T_{\rm in}$ and $T_{\rm sky}$, then the ratio is

\begin{equation}
  \frac{\sigma(T_{\rm sky})}{\sigma(T_{\rm in})} = \frac{ V_{\rm out}(T_{\rm sky})}
   {V_{\rm out}(T_{\rm in})}\times
   \frac{1+b\, G_0(T_{\rm in}+T_{\rm noise})}{1+b\, G_0 (T_{\rm sky}+T_{\rm noise})}.
   \label{eq:ratio_uncalibrated_white_noise}
\end{equation}

If we call $\rho$ the ratio $\frac{\sigma(T_{\rm sky})}{\sigma(T_{\rm in})} $ and use Eq.~(\ref{eq:nonlinear_response})
to put in explicit form the ratio of output voltages in Eq.~(\ref{eq:ratio_uncalibrated_white_noise}) so that
$\sigma(T_{\rm sky}) = \rho\times \sigma(T_{\rm in})$, we have

\begin{equation}
 \rho = \frac{
  T_{\rm sky}+T_{\rm noise}}{T_{\rm in}+T_{\rm noise} }\times
   \left[\frac{1+ b\, G_0
   (T_{\rm in}+T_{\rm noise})}{1+b\, G_0
   (T_{\rm sky}+T_{\rm noise})}\right]^2.
   \label{eq:ratio_uncalibrated_white_noise_1}
\end{equation}

\subsection{Step 4---calibrate extrapolated noise}

From Eqs.~(\ref{eq:white_noise}) and (\ref{eq:nonlinear_response}) we obtain 

\begin{equation}
 \sigma = \frac{G_0}{\left[1+b\, G_0(T_{\rm sky}+T_{\rm noise})\right]^2}\times 2\frac{T_{\rm sky}+T_{\rm
noise}}{\sqrt{\beta}}.
 \label{eq:tilde_wn_final}
\end{equation}

If we call $\sigma^{\rm cal}$ the calibrated noise extrapolated at the sky temperature and consider that,
by definition, $\sigma^{\rm cal} = 2\frac{T_{\rm sky}+T_{\rm noise}}{\sqrt{\beta}}$, the previous equation gives

\begin{equation}
  \sigma^{\rm cal} =  \frac{\left[1+b\, G_0(T_{\rm sky}+T_{\rm noise})\right]^2}{G_0} \sigma.
\end{equation}

\section{Weighted noise averaging}
\label{app:weighted_noise_average}

   According to the LFI receiver design the output from each radiometer results from the 
   combination of signals coming from two corresponding detector diodes. Consider
   two differenced and calibrated
   datastreams coming from two detectors of a radiometer leg, $d_1(t)$ and $d_2(t)$. The simplest 
   way to combine the two outputs is to take a straight average, i.e.,

   \begin{equation}
      d(t) = \frac{d_1(t)+d_2(t)}{2},
      \label{eq:straight_diode_averaging}
   \end{equation}
   so that the white noise level of the differenced datastream is given by 
   $\sigma_{d(t)} = \sqrt{\sigma_{d_1(t)}^2 + \sigma_{d_2(t)}^2}$.

   This approach, however, is not optimal
   in cases where the two noise levels are unbalanced, so that the noise of the averaged datastream is dominated
   by the noisier channel.

   An alternative to Eq.~(\ref{eq:straight_diode_averaging}) is given by a weighting average in which weights
   are represented by the inverse of the noise levels of the two diode datasteams, i.e.,

   \begin{equation}
      d(t) = \frac{w_1 d_1(t) + w_2 d_2(t)}{w_1 + w_2},
      \label{eq:weighted_diode_averaging}
   \end{equation}
   or, more generally, in the case where we want to average more than two datastreams together,
    
   \begin{equation}
        d(t) = \frac{\sum_{j=1}^N w_j d_j(t)}{\sum_{j=1}^N w_j}.
        \label{eq:weighted_diode_averaging_general}
   \end{equation}
  For noise-weighted averaging, we choose the weights as $w_j = \sigma_{d_j(t)}^{-2}$ so that the white noise of the
differenced datastream is given by:

   \begin{equation}
      \sigma_{d(t)} = \left(\sum_{j=1}^N\sigma_{d_j(t)}^{-2}\right)^{-1/2}.
      \label{eq:weighted_white_noise}
   \end{equation}

   \section{Thermal susceptibility scientific requirement}
   \label{app:susceptibility_scientific_requirement}
      
        Temperature fluctuations in the LFI focal plane arise primarily from variations in the sorption  cooler system
driven by the cycles of the six cooler compressors that ``pump''\footnote{The sorption cooler does not use mechanical
compressors to generate a high pressure flow, but a process of absorption-desorption of hydrogen into six hydride beds,
the ``compressors'' controlled by a temperature modulation of the beds themselves.} the hydrogen in the high pressure
piping line to the cooler cold-end. These fluctuations show a frequency spectrum dominated by a period of $\sim$1~hour,
corresponding to the global warm-up/cool-down cycle of the six compressors.

        An active PID temperature stabilisation assembly at the interface between the cooler cold-end and the focal
plane achieves stabilities of the order of 80$-$100\,mK peak-to-peak with a frequency spectrum dominated by the single
compressor frequency ($\sim 1$\,mHz) and the frequency of the whole assembly ($\sim 0.2$\,mHz).

        These fluctuations propagate through the focal plane mechanical structure, so that the actual temperature
instabilities at the level of the feed-amplifier systems (the term $\Delta T_{\rm phys}$ in
Eq.~(\ref{eq:susceptibility_transfer_function})) are significantly damped. The LFI thermal model \citep{2009_LFI_cal_T3}
shows that the fluctuations at the front-end modules are at the level of $\lesssim 10$~mK and dominated by the
``slowest'' components (i.e., those with frequencies $\lesssim 10^{-2}$~Hz).
        
        If we take into account that slow fluctuations in the antenna temperature time stream are further damped by a
factor $\sim 10^3$ by the scanning strategy and map-making \citep{mennella02}, we can easily see from
Eq.~(\ref{eq:susceptibility_transfer_function}) that a receiver susceptibility $f_{\rm trans} \lesssim 0.1$ is required
to maintain the final peak-to-peak error per pixel $\lesssim 1 \mu$K.

\section{Front-end temperature susceptibility parameters}
\label{app:susceptibility parameters}

    Temperature susceptibility parameters are summarised in Table~\ref{tab:THF_susceptilility_results}.

   \begin{table}[h!]
      \caption{
         Gain and noise temperature susceptibilities to front-end temperature fluctuations
         measured during the RCA calibration campaign.
      }
      \label{tab:THF_susceptilility_results}
      \begin{center}
         \begin{tabular}{l c c c c}
                     & \multicolumn{4}{c}{$\partial G / \partial T_{\rm phys}$(dB/K)}    \\
            \cline{2-5}
                     & M-00  & M-01  & S-10  & S-11\\
            \hline
            \hline
            LFI18       &-0.05  &-0.05  &-0.05  &-0.05\\
            LFI19       &-0.05  &-0.04  &-0.02  &-0.03\\
            LFI20       &-0.05  &-0.04  &-0.03  &-0.04\\
            LFI21       &-0.07  &-0.07  &-0.07  &-0.20\\
            LFI22       &-0.21  &-0.15  &-0.18  &-0.13\\
            LFI23       &-0.03  &-0.05  &-0.05  &-0.05\\
            LFI24       &-0.08  &-0.06  &-0.08  &-0.08\\
            LFI25       &-0.02  &-0.02  &-0.04  &-0.05\\
            LFI26       &-0.01  &-0.03  &-0.01  &-0.01\\
            LFI27       &-0.06  &-0.05  &-0.04  &-0.01\\
            LFI28       &-0.03  &-0.07  &-0.14  &-0.13\\
            \hline
            \multicolumn{5}{c}{ }
         \end{tabular}
\vspace{2cm}
         \begin{tabular}{l c c c c}
                     & \multicolumn{4}{c}{$\partial T_{\rm noise} / \partial T_{\rm phys}$(K/K)}    \\
            \cline{2-5}
                     & M-00  & M-01  & S-10  & S-11\\
            \hline
            \hline
               LFI18    &0.47   &0.49   &0.38   &0.42\\
               LFI19    &0.36   &0.33   &0.40   &0.37\\
               LFI20    &0.25   &0.23   &0.30   &0.25\\
               LFI21    &0.15   &0.15   &0.18   &0.30\\
               LFI22    &0.10   &0.10   &0.10   &0.10\\
               LFI23    &0.10   &0.16   &0.17   &0.16\\
               LFI24    &0.40   &0.41   &0.10   &0.43\\
               LFI25    &0.12   &0.10   &0.25   &0.08\\
               LFI26    &0.70   &0.70   &0.47   &0.50\\
               LFI27    &0.81   &0.45   &0.58   &0.34\\
               LFI28    &0.15   &0.15   &0.10   &0.33\\
            \hline
         \end{tabular}
      \end{center}
   \end{table}

\bibliographystyle{aa}
\bibliography{references}
\end{document}